\documentclass[sigconf]{aamas}

\usepackage{amsmath,mathtools,bm}
\usepackage{booktabs}
\usepackage{microtype}
\usepackage{xcolor}
\newcommand{\rev}[1]{#1}
\usepackage{graphicx}
\usepackage{tikz}
\usepackage{enumitem}
\usepackage{multirow}
\usepackage{array}
\usetikzlibrary{arrows.meta,positioning,fit,calc}
\usepackage{wrapfig}
\usepackage{amsfonts}
\usepackage{amsthm}
\usepackage{algorithm}
\usepackage{algorithmic}
\usepackage{tocloft}

\newcommand{\AUC}{\mathrm{AUC}}
\newcommand{\Gap}{\mathrm{Gap}}
\newcommand{\Simplex}{\Delta}
\newcommand{\R}{\mathbb{R}}

\newtheorem{theorem}{Theorem}
\newtheorem{proposition}{Proposition}
\newtheorem{corollary}{Corollary}

\setcopyright{none}
\renewcommand\footnotetextcopyrightpermission[1]{}
\renewcommand{\contentsname}{APPENDIX}

\renewcommand{\cfttoctitlefont}{\Large\bfseries}
\renewcommand{\cftaftertoctitle}{\hfill}

\setcopyright{none}
\acmConference[]{}{}{}
\acmBooktitle{}
\makeatletter
\AtBeginDocument{%
  \renewcommand{\tableofcontents}{%
    \par
    \begingroup
      \parindent\z@ \parskip\cftparskip
      {\interlinepenalty\@M
       {\cfttoctitlefont\contentsname}{\cftaftertoctitle}\par\nobreak
       \vskip\cftaftertoctitleskip
       \@afterheading}%
      \makeatletter
      \@input{\jobname.toc}%
      \if@filesw
        \expandafter\newwrite\csname tf@toc\endcsname
        \immediate\openout\csname tf@toc\endcsname\jobname.toc\relax
      \fi
      \global\@nobreakfalse
    \endgroup}}
\makeatother

\title{The Right Solver at the Right Time:\texorpdfstring{\\}{} Amortised Nash Equilibrium Computation}

\author{Yaqi Sun}
\authornote{Equal contribution.}
\affiliation{
  \institution{Queen Mary University of London}
  \city{London}
  \country{United Kingdom}}

\author{Julian Ma}
\authornotemark[1]
\affiliation{
  \institution{University College London}
  \city{London}
  \country{United Kingdom}}

\author{Jing Peng}
\affiliation{
  \institution{Amazon}
  \country{United States}}

\author{Claudia Garetto}
\affiliation{
  \institution{Queen Mary University of London}
  \city{London}
  \country{United Kingdom}}

\author{David Mguni}
\authornote{Corresponding author: \href{mailto:d.mguni@qmul.ac.uk}{d.mguni@qmul.ac.uk}}
\affiliation{
  \institution{Queen Mary University of London}
  \city{London}
  \country{United Kingdom}}

\begin{document}
\addtocontents{toc}{\protect\setcounter{tocdepth}{0}}


\begin{abstract}
Efficient Nash equilibrium (NE) computation underpins applications in multi-agent
learning, economics, and auctions, yet current pipelines typically make a static
computational choice: they initialise a solve and then apply a fixed solver, or
select a solver once for the game. We show that this misses structure at two
levels: the most suitable solver can vary across games, and its suitability can
change as optimisation unfolds. A method that is effective early may later
stagnate, oscillate or cycle, while another becomes better suited to the evolving
optimisation state; moreover, the quality of a warm start depends on the solver
policy that follows it. We address these limitations with a three-layer framework
for \emph{amortised NE computation}: a frozen behavioural representation provides
solver-relevant game context, a controller-aware equivariant initialiser learns
where to start, and a state-dependent controller selects the solver mechanism at
the beginning of the solve and revises that choice as the realised trajectory
evolves. The controller switches among persistent primitives capturing core
mechanisms of modern iterative NE solvers, while the initialiser is refined
against downstream controller value. Together, these components learn where
equilibrium search should begin, which solver should govern it initially, and
when that solver should change. On held-out
$3\times3$, $4\times4$ and $5\times5$ general-sum games generated after model selection, our method reduces
normalised exploitability AUC at $C=200$ operator calls by \rev{43.8\%, 26.5\% and 32.6\%}, respectively, relative to the original warm start with fixed Optimistic.
Dynamic control remains beneficial across all three game sizes. These results suggest a broader shift from designing a single NE solver to learning computational policies that orchestrate complementary solver mechanisms across the course of equilibrium search.
\end{abstract}
\maketitle

\section{Introduction}
\label{sec:introduction}

Nash equilibrium (NE) computation is a recurring primitive in multi-agent systems and
algorithmic game theory, with applications spanning mechanism design, auctions
and learning among strategic agents~\cite{nisan2007algorithmic,li2024survey}.
In many such applications the computational task is not a single isolated game
but a sequence of related games, so repeatedly solving each instance from
scratch discards structure shared across problems~\cite{harris2023metalearning}.
There is also no reason to expect one iterative mechanism to be uniformly best.
First-order game dynamics can exhibit rotational or recurrent behaviour
~\cite{daskalakis2018limit,balduzzi2018mechanics}, while optimistic,
extra-gradient and related corrections alter stability in important game
classes~\cite{mertikopoulos2019optimistic,gidel2019variational}. More broadly,
empirical and theoretical studies of finite normal-form games show substantial
variation across equilibrium-solving methods and game families
~\cite{li2024survey}. This creates two opportunities for reuse: experience can
inform both where equilibrium search should begin and which computational
mechanism is appropriate as the solve develops.

This leads to a different question from conventional algorithm selection, which
maps a problem instance to a single algorithm or portfolio choice
~\cite{rice1976algorithm}. Can previous experience instead be turned into a
\emph{computational policy} that decides both where a new solve should start and
how it should proceed? The two decisions are coupled. Warm-starting can
substantially accelerate equilibrium computation~\cite{brown2016warmstart,duan2023learnable},
but the best starting point depends on the refinement procedure that follows it;
conversely, the best refinement mechanism can depend on the state created by the
initialiser and by earlier solver updates.

\begin{figure*}[!t]
  \centering
  \includegraphics[width=0.6\textwidth]{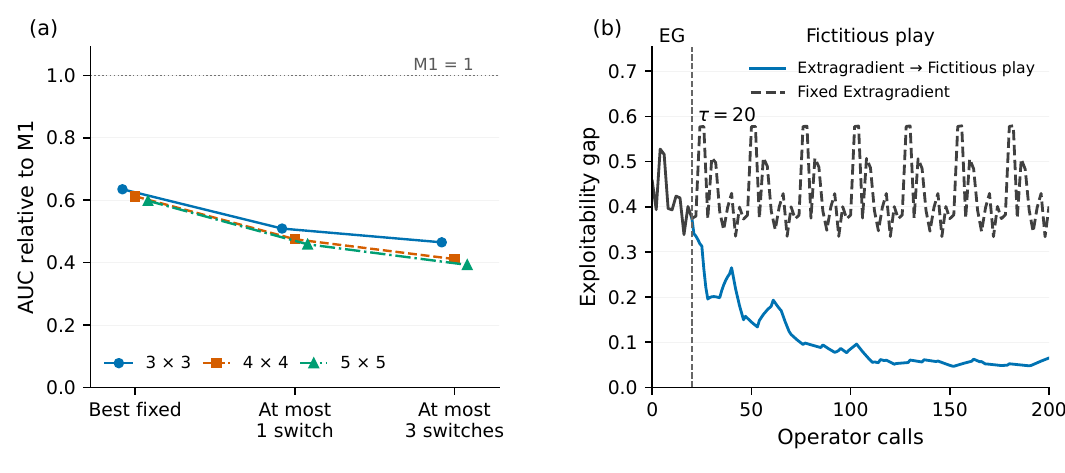}
  \caption{Solver choice is phase-dependent.
  (a)~AUC of the exploitability gap over $C=200$ operator calls, relative to M1 (lower is better), for the per-game oracle's best fixed solver, the best schedule with at most one switch, and the best schedule with at most three switches.
  (b)~Game~362 ($5\times5$, warm seed~1). Both runs use extragradient (EG) and are identical up to $\tau=20$. Continuing EG leads to oscillatory behaviour, whereas switching to fictitious play at $\tau$ reduces the AUC by 68.66\%.}
  \label{fig:phase-dependent}
\end{figure*}

We address this problem with the framework in
Section~\ref{sec:framework}. A frozen behavioural representation $\hat z(G)$
provides game-level context. An action-equivariant meta-initialiser maps the game
to a starting strategy, while a state-dependent controller observes the evolving
solve and chooses among a fixed library of persistent equilibrium-solving
primitives. The library contains recurring mechanisms from iterative equilibrium
computation: simultaneous gradients, optimism/extrapolation, mirror-style
updates, damped projected updates, anchoring and fictitious play
~\cite{li2024survey}. The controller can therefore change mechanism when the
learning dynamics change, for example leaving an update rule that begins to
oscillate in favour of one that makes more stable progress (Figure~\ref{fig:phase-dependent}b).

The starting point and the controller cannot be optimised independently without
care. A warm start tuned for one fixed solver may be poorly matched to a dynamic
controller. Full end-to-end differentiation is also awkward because the
controller makes discrete switching decisions. Section~\ref{sec:value-guided-init}
therefore introduces a lightweight bridge between the two: after training the
controller, we fit a differentiable model of downstream controller performance
around the current warm starts and use its gradient to refine only the
initialiser. This makes the initial point controller aware without backpropagating
through the discrete solver trajectory.

The second part of the framework is sequential by construction. Choosing a
primitive at state $\theta_t$ changes $\theta_{t+1}$ and hence the states from
which future choices are made. A locally attractive update can therefore be a
poor long-run decision, while an apparently weaker step may move the trajectory
into a region from which equilibrium is reached faster. We formulate this as a
finite-horizon switching-control problem in Sections~\ref{sec:problem} and
\ref{sec:controller}, with a small cost for changing solver regime and exact
payoff/operator-call accounting. Section~\ref{sec:theory} gives the corresponding
budgeted Bellman recursion and its finite-horizon optimality result.

Our experiments in Section~\ref{sec:experiments} are organised around five
questions: whether a strong start removes the need for control, whether feedback
from the trajectory is genuinely useful, whether controller-aware initialisation
and adaptive control add up, whether the gains persist when the compute
horizon is extended, and whether the controller transfers to larger and unseen game sizes. The results separate these effects clearly: a strong start helps most on
typical games, while adaptive control is especially valuable on harder games
where a fixed solver stalls or oscillates. On held-out games,
combining the two reduces normalised AUC by \rev{26.5--43.8\%} relative to the
original warm start with fixed Optimistic. Appendices G, J and K report the initialiser analysis, controller ablations, restricted switching oracles, representation diagnostics, size transfer, elapsed time and portfolio ablations.

\paragraph{Contributions.}
Our contributions address key limitations of current equilibrium-solving
pipelines:

\begin{enumerate}[leftmargin=*,itemsep=2pt,topsep=3pt]

    \item \textbf{Phase-dependent equilibrium solving.}
    We formulate repeated NE computation as a finite-budget
    switching problem in which solver choice can adapt as the trajectory evolves
    (Sections~\ref{sec:problem}--\ref{sec:controller}). The resulting Bellman
    formulation characterises an optimal switching policy, while trajectory-aware
    control improves over static per-game solver selection by \(11.7\)--\(14.6\%\).

    \item \textbf{Controller-aware initialisation.}
    Because the value of a warm start depends on the solver policy that follows
    it, we refine the action-equivariant initialiser using a differentiable local
    model of downstream controller value, without differentiating through discrete
    switches (Section~\ref{sec:value-guided-init}). This improves the full pipeline
    while retaining the benefit of later adaptive control.

    \item \textbf{Persistent solver control over the full trajectory.}
    We combine frozen solver-relevant game context with realised trajectory state
    to switch among persistent optimisation primitives rather than selecting a
    solver only once (Sections~\ref{sec:framework} and \ref{sec:controller}).
    Restricted exact searches show that later switching provides substantial
    headroom beyond choosing the best initial primitive.

    \item \textbf{Matched-compute evidence across starts and horizons.}
    Under operator-call accounting, we show that improved initialisation and
    dynamic control make complementary contributions, remain valuable on hard
    games, and transfer beyond the \(C=200\) training horizon; a controller
    trained on 3--100 actions also transfers to held-out sizes
    (Sections~\ref{sec:experiments} and~\ref{sec:scale-transfer}; Appendices G and J).

\end{enumerate}

\section{Related Work}
\label{sec:related}

\paragraph{Iterative equilibrium solvers and game dynamics.}
A central difficulty in learning in games is that optimisation dynamics can be
qualitatively different from those of single-objective optimisation. Simultaneous
gradient descent/ascent can exhibit recurrent or rotational behaviour even in
simple adversarial problems, while optimistic and extra-gradient mechanisms can
change the stability of the resulting dynamics~\cite{daskalakis2018limit,mertikopoulos2019optimistic}.
Variational-inequality formulations motivate extrapolation and averaging as
principled ways to control adversarial optimisation~\cite{gidel2019variational},
and the decomposition of differentiable games into potential and Hamiltonian
components gives another account of why rotation can obstruct convergence and
why different corrections can be useful in different regimes~\cite{balduzzi2018mechanics}.
For fictitious play, decomposing a game into zero-sum and potential components
yields convergence conditions for games that interpolate between the two
classes~\cite{chen2022fpdecomposition}.
These works design or analyse particular update mechanisms. Our concern is
orthogonal: when several such mechanisms are available, which should govern the
current phase of a previously unseen solve?

\textit{Warm starts and amortised game solving.}
Several strands of work exploit related games to reduce repeated equilibrium
computation. Brown and Sandholm develop a principled strategy-based warm start
for CFR in extensive-form games~\cite{brown2016warmstart}. Harris et
al.~\cite{harris2023metalearning} formulate meta-learning for equilibrium
finding and obtain rates that adapt to similarity across sequences of games.
Duan et al.~\cite{duan2023learnable} study learned Nash approximators and show
that their predictions can warm-start classical solvers. NfgTransformer instead
builds permutation-equivariant representations of normal-form games and applies
them to equilibrium prediction and related game-theoretic tasks~\cite{liu2024nfgtransformer}.
These approaches motivate learning \emph{where} to begin. Our experiments use
strong learned initialisers as comparators, but ask a complementary question:
after a starting point has been chosen, can the refinement mechanism itself
adapt online to the realised trajectory?

\rev{NES~\cite{marris2022nes} learns mappings from games to CE/CCE and obtains NE from CCE marginals in two-player constant-sum games. RENES~\cite{wang2024renes} learns payoff modifications that improve the strategies a downstream solver returns for the original game. Our controller instead allocates solver primitives along a single trajectory on the original game.}

\textit{Algorithm selection and learned optimisation.}
The classical algorithm-selection problem asks how features of a problem can be
used to choose among algorithms with different performance profiles
~\cite{rice1976algorithm}. Learned-optimiser methods go further by learning the
update rule itself from a distribution of optimisation tasks
~\cite{andrychowicz2016learning}. Our setting lies between these views. We retain
interpretable, stateful equilibrium-solving primitives rather than replacing
them with a monolithic learned optimiser, but the choice among them is not a
one-shot decision. It is a sequential policy: the controller may revise its
choice after observing the state produced by earlier solver updates. Static per-game solver selection is therefore a natural comparator for isolating the value of trajectory feedback.

\rev{Dynamic algorithm configuration learns policies that adapt algorithm parameters during a run across problem instances~\cite{biedenkapp2020dac}, and Schr\"oder et al.~\cite{schroder2022switching} study switching between black-box optimisers with warm-starting policies. LICRA~\cite{mguni2023licra} combines reinforcement learning with impulse control to learn when to act and which action to take under action costs and budget constraints. We apply this kind of in-run adaptation to equilibrium computation: the controller repeatedly chooses among persistent NE primitives under a payoff-call budget, and the initialiser is refined for that controller.}

\textit{Behavioural representations and solver synthesis.}
A solver-induced behavioural representation provides the game-level context used in our framework. Related work maps games into a solver-aligned behavioural geometry and uses this representation for structure-aware solver synthesis through continuous combinations of primitive mechanisms~\cite{sun2026geometrygames}. We take this representation as a fixed input rather than as an object of learning or a contribution of the present paper. Our focus is instead on how such game-level context can be made computationally actionable: first through controller-aware initialisation, and then through trajectory-dependent switching among persistent solver mechanisms as the solve unfolds.

Taken together, these strands leave a gap between \emph{where to start} and
\emph{how to continue}. Warm-start and meta-learning methods transfer
information across games but typically leave the downstream solver fixed;
classical algorithm selection chooses among methods but ordinarily makes the
choice at the instance level; learned optimisers can adapt updates, but at the
cost of replacing the solver mechanism itself. Our framework keeps established
solver mechanisms intact and uses feedback to allocate them over a trajectory,
while separately learning a starting point that is valuable under that
controller. This separation is what the experiments in
Section~\ref{sec:experiments} test directly.

\section{Problem Formulation}
\label{sec:problem}

\paragraph{Games and exploitability.}
We consider two-player normal-form games $G=(A,B)$ with
$A,B\in\R^{n\times m}$, row strategy $x\in\Simplex^n$, column strategy
$y\in\Simplex^m$, and joint strategy $\theta=(x,y)$. Equilibrium quality is
measured by the Nash gap
\begin{equation}
\Gap_G(\theta)
=
\max_{x'\in\Simplex^n}x'^\top Ay-x^\top Ay
+
\max_{y'\in\Simplex^m}x^\top By'-x^\top By,
\label{eq:gap}
\end{equation}
which is non-negative and vanishes at a NE.

Payoffs are normalised by subtracting each player’s mean payoff and dividing both payoff matrices by their shared maximum absolute centred entry (floored at \(10^{-8}\)). This places payoffs in \([-1,1]\), giving Nash gaps a common payoff scale across games and sizes.

\paragraph{Stateful solver primitives.}
Let $\mathcal I=\{1,\ldots,M\}$ index a fixed library of iterative
solver mechanisms. Primitive $j$ is stateful: from strategy $\theta_t$ and
internal memory $m_t^{(j)}$ it produces
\begin{equation}
\mathcal T_j(G,\theta_t,m_t^{(j)})
=
\bigl(\theta_{t+1},\widetilde m_{t+1}^{(j)},c_t^{(j)}\bigr),
\label{eq:primitive-transition}
\end{equation}
where $c_t^{(j)}$ is the number of payoff/operator evaluations used by the
update. The memory may contain previous-gradient information, empirical play,
running averages or anchors. When a primitive becomes inactive its memory is
retained; switching solver does not silently restart its algorithmic history.

\paragraph{Compute clock and control.}
Nominal iterations are not comparable across primitives, so cumulative compute
is measured by operator calls. Control decisions are taken on a fixed grid, every $\Delta$
operator calls. We index decision points by $t$, write $C_t$ for the compute
used before decision $t$ ($C_0=0$), and let $\rho_t=C_t/C_{\max}$.
Before decision $t$, let $i_t$ be the active primitive; $i_0$ is a fixed
reference primitive. A gate samples $g_t\in\{\textsc{Stay},\textsc{Switch}\}$
and, conditional on switching, a router samples
$j_t\in\mathcal I\setminus\{i_t\}$. The primitive executed in the next segment is
\begin{equation}
i_t^+=
\begin{cases}
i_t, & g_t=\textsc{Stay},\\
j_t, & g_t=\textsc{Switch},
\end{cases}
\qquad i_{t+1}=i_t^+.
\label{eq:active-after-decision}
\end{equation}
Primitive $i_t^+$ then performs as many complete updates as fit in the next
$\Delta$ calls, or in the remaining budget; $\Delta C_t=C_{t+1}-C_t$ denotes the
calls actually used. Only the executed primitive changes the realised strategy trajectory.

\paragraph{Objective.}
We care about how quickly exploitability is reduced throughout the solve, not
only about the terminal iterate. For a controller $\pi$, we therefore minimise
\begin{equation}
J(\pi)=
\mathbb E_{G}\,\mathbb E_\pi\!\left[
\frac{1}{C_{\max}}\int_0^{C_{\max}}\Gap_G(\theta(c))\,dc
+\lambda_{\mathrm{sw}}N_{\mathrm{sw}}
\right],
\label{eq:controller-objective}
\end{equation}
where $N_{\mathrm{sw}}=\sum_t\mathbf 1[g_t=\textsc{Switch}]$. The switching
penalty discourages unnecessary regime changes; solver cost itself is already
accounted for on the operator-call axis. For a realised transition consuming
$\Delta C_t=C_{t+1}-C_t$, the corresponding dense reward is
\begin{align}
r_t={}-\frac{\Delta C_t}{C_{\max}}
\frac{\Gap_G(\theta_t)+\Gap_G(\theta_{t+1})}{2}-\lambda_{\mathrm{sw}}\mathbf 1[g_t=\textsc{Switch}].
\label{eq:reward}
\end{align}
When a multi-call primitive exposes intermediate iterates, we integrate on the
finer operator-call grid.

For a target quality $\varepsilon$, we also report the hitting cost
$\tau_\varepsilon=\inf\{C\le C_{\max}:\Gap_G(\theta(C))\le\varepsilon\}$ and the
success probability $\Pr(\tau_\varepsilon\le C_{\max})$. Failures are retained
as censored/unsuccessful episodes rather than removed from the comparison.

\section{Amortised Equilibrium Computation}
\label{sec:framework}

\subsection{Overview}
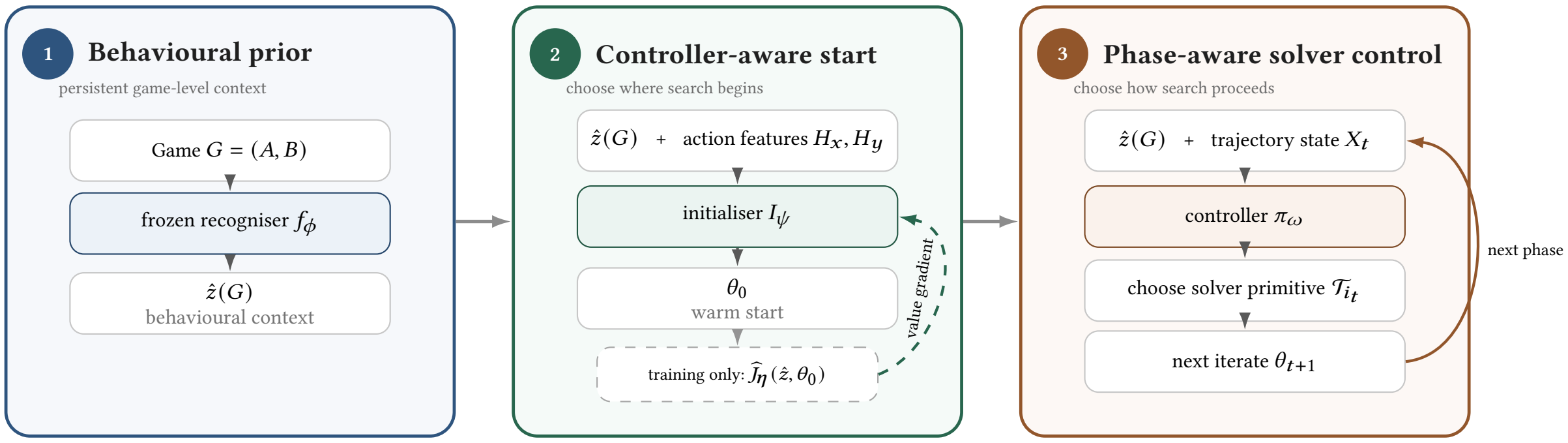
\begin{figure*}[t]
\centering
\begin{tikzpicture}[
  font=\sffamily,
  panel/.style={draw=#1!70!black, fill=#1!5, rounded corners=7pt,
                line width=0.9pt, minimum width=4.55cm, minimum height=4.35cm},
  title/.style={font=\bfseries\small, text=black!88},
  tag/.style={circle, fill=#1!70!black, text=white, font=\bfseries\scriptsize,
              minimum size=5.2mm, inner sep=0pt},
  box/.style={draw=black!22, fill=white, rounded corners=4pt, align=center,
              minimum width=3.25cm, minimum height=0.62cm, inner sep=2.5pt,
              font=\scriptsize},
  keybox/.style={box, draw=#1!60!black, fill=#1!10},
  ghost/.style={draw=black!35, dashed, fill=white, rounded corners=4pt,
                align=center, minimum width=2.85cm, minimum height=0.55cm,
                inner sep=2pt, font=\tiny},
  arr/.style={-{Latex[length=1.8mm,width=1.2mm]}, line width=0.8pt,
              draw=black!65},
  link/.style={-{Latex[length=2.0mm,width=1.3mm]}, line width=1.0pt,
               draw=black!45}
]
\definecolor{StageBlue}{RGB}{66,120,180}
\definecolor{StageGreen}{RGB}{57,145,112}
\definecolor{StageOrange}{RGB}{208,123,61}

\node[panel=StageBlue]   (p1) at (0,0) {};
\node[panel=StageGreen]  (p2) at (5.15,0) {};
\node[panel=StageOrange] (p3) at (10.30,0) {};

\node[tag=StageBlue] at (-1.85,1.70) {1};
\node[title,anchor=west] at (-1.55,1.70) {Behavioural prior};
\node[font=\tiny,text=black!58,anchor=west] at (-1.85,1.34)
{persistent game-level context};
\node[box] (game) at (0,0.72) {Game $G=(A,B)$};
\node[keybox=StageBlue] (atlas) at (0,-0.02) {frozen recogniser $f_\phi$};
\node[box] (z) at (0,-0.83) {$\hat z(G)$\\\textcolor{black!55}{behavioural context}};
\draw[arr] (game) -- (atlas);
\draw[arr] (atlas) -- (z);

\node[tag=StageGreen] at (3.30,1.70) {2};
\node[title,anchor=west] at (3.60,1.70) {Controller-aware start};
\node[font=\tiny,text=black!58,anchor=west] at (3.30,1.34)
{choose where search begins};
\node[box] (feat) at (5.15,0.82) {$\hat z(G)$ \; + \; action features $H_x,H_y$};
\node[keybox=StageGreen] (init) at (5.15,0.05) {initialiser $I_\psi$};
\node[box] (theta) at (5.15,-0.78) {$\theta_0$\\\textcolor{black!55}{warm start}};
\node[ghost] (value) at (5.15,-1.55)
{training only: $\widehat J_\eta(\hat z,\theta_0)$};
\draw[arr] (feat) -- (init);
\draw[arr] (init) -- (theta);
\draw[arr,dashed,draw=black!42] (theta) -- (value);
\draw[arr,dashed,draw=StageGreen!70!black]
(value.east) .. controls +(0.65,0.25) and +(0.75,-0.15) ..
node[font=\tiny,sloped,above]{value gradient} (init.east);

\node[tag=StageOrange] at (8.45,1.70) {3};
\node[title,anchor=west] at (8.75,1.70) {Phase-aware solver control};
\node[font=\tiny,text=black!58,anchor=west] at (8.45,1.34)
{choose how search proceeds};
\node[box] (state) at (10.30,0.82) {$\hat z(G)$ \; + \; trajectory state $X_t$};
\node[keybox=StageOrange] (ctrl) at (10.30,0.05) {controller $\pi_\omega$};
\node[box] (prim) at (10.30,-0.70) {choose solver primitive $\mathcal T_{i_t}$};
\node[box] (next) at (10.30,-1.42) {next iterate $\theta_{t+1}$};
\draw[arr] (state) -- (ctrl);
\draw[arr] (ctrl) -- (prim);
\draw[arr] (prim) -- (next);
\draw[arr,draw=StageOrange!70!black]
(next.east) .. controls +(0.95,0.10) and +(0.95,-0.10) ..
node[font=\tiny,right]{next phase} (state.east);

\draw[link] (p1.east) -- (p2.west);
\draw[link] (p2.east) -- (p3.west);
\end{tikzpicture}
\caption{\textbf{Amortised equilibrium computation.} The framework separates
three roles: a frozen behavioural prior supplies game-level context, a
controller-aware initialiser chooses where equilibrium search begins, and a
dynamic controller chooses which solver mechanism should govern the current
phase of the solve. The local value model in the centre panel is used only to
refine the initialiser during training; deployment follows the solid path.}
\label{fig:pipeline}
\end{figure*}

The framework amortises two decisions that are usually made independently:
where a solve begins and which computational mechanism should refine it. Given a
game $G$, the frozen recogniser produces $\hat z(G)$, the initialiser produces
$\theta_0=I_\psi(G,\hat z(G))$, and the controller chooses solver primitives from an observation $o_t$ that combines $\hat z(G)$ with a summary of the solve so far (Section~\ref{sec:controller}). Figure~\ref{fig:pipeline}
shows the resulting flow. The representation supplies persistent context about
the game; the dynamic state tells the controller what has happened so far. This
separation allows two games with similar global structure to require different
solver choices once their trajectories diverge, and allows the same game to
favour different mechanisms at different stages.

\subsection{Frozen behavioural prior}
The game context is the fixed representation $\hat z(G)=f_\phi(A,B)\in\R^d$,
obtained from solver-induced behavioural distances through a low-dimensional
atlas and a symmetry-aware payoff encoder~\cite{sun2026geometrygames}; $\phi$ is
frozen throughout. The representation supplies a compact game-level prior to
both the initialiser and the controller, while information about the ongoing
solve reaches the controller separately.

\rev{The underlying atlas uses 20,000 $3\times3$ training games and 31 solver configurations: five step sizes for each of six mechanisms and one fictitious-play configuration, each run for 200 calls. A payoff recogniser predicts atlas coordinates and is frozen before controller training.}

\subsection{Controller-aware meta-initialisation: where to start}
\label{sec:value-guided-init}
The symmetry-aware encoder exposes row- and column-action embeddings
$H_x(G)=(h^x_1,\ldots,h^x_n)$ and $H_y(G)=(h^y_1,\ldots,h^y_m)$. The base
initialiser assigns action logits and normalises them within each player,
\begin{align}
a_i&=q^x_\psi(h^x_i,\hat z(G)), &
 x_{0,i}&=\frac{e^{a_i}}{\sum_k e^{a_k}},\\
b_j&=q^y_\psi(h^y_j,\hat z(G)), &
 y_{0,j}&=\frac{e^{b_j}}{\sum_\ell e^{b_\ell}}.
\label{eq:init}
\end{align}
The action embeddings are necessary as $\hat z$ is invariant to action
relabeling whereas the mixed strategy must transform equivariantly.

A starting point is only as good as the computation that follows it: a warm
start trained without regard to the downstream solver need not be the best state
from which a switching controller operates. We therefore refine the initialiser
using the value of a trained, frozen controller $\pi_\omega$, without
differentiating through its discrete switches. Around each base warm start, we generate local perturbations $\{\widetilde\theta_0^{(k)}\}_{k=0}^{K}$ by adding independent Gaussian noise to each player’s pre-softmax logits and then applying softmax. Specifically, $\widetilde\ell=\ell+\sigma\,\mathrm{std}(\ell)\varepsilon$, where $\varepsilon\sim\mathcal N(0,I)$, the standard deviation is computed across that player’s logits, and $\sigma\in\{0.25,0.5,1.0\}$ is assigned cyclically across games and perturbations. We use $K=8$. The target for perturbation $k$ is
\begin{equation}
J_{\pi_\omega}(G,\widetilde\theta_0^{(k)})
=\mathbb E_{\pi_\omega}\!\left[\AUC_{C_{\max}}(G,\pi_\omega\mid\widetilde\theta_0^{(k)})
+\lambda_{\mathrm{sw}}N_{\mathrm{sw}}\right].
\label{eq:init-value-target}
\end{equation}
A local value model $\widehat J_\eta(\hat z(G),E_\theta(\widetilde\theta_0))$ is
fitted to these targets on training games, where $E_\theta$ is the strategy
encoder also used by the controller (Section~\ref{sec:controller}). Sampling
nearby starts forces \(\widehat J_\eta\) to capture how the value changes with $\theta_0$, rather than only the game difficulty.

Let $\theta_\psi(G)=I_\psi(G,\hat z(G))$. After fitting the surrogate, both
$\pi_\omega$ and $\widehat J_\eta$ are frozen and only the initialiser is
updated:
\begin{align*}
\mathcal L_{\mathrm{VG}}(\psi)
={}&\mathbb E_G\!\left[
\widehat J_\eta\bigl(\hat z(G),E_\theta(\theta_\psi(G))\bigr)
\right]
+\beta D_{\mathrm{KL}}\!\left(
\theta_\psi(G)\,\middle\|\,\theta_{\psi_0}(G)
\right).
\end{align*}
For $\theta=(x,y)$ we use the sum of the two player-wise KL divergences. The KL
term keeps the update near the region covered by the local value model. Crucially,
$\widehat J_\eta$ is only a training device: model selection and final evaluation
always use real controller rollouts from the refined warm start.

\subsection{Phase-dependent solver control}
\label{sec:controller}
The controller needs both persistent game information and a summary of the
realised solve. We write its augmented state as
\begin{equation}
X_t=(\hat z(G),\theta_t,i_t,\rho_t,\mathcal M_t,h_t),
\label{eq:augmented-state}
\end{equation}
where $\mathcal M_t=\{m_t^{(1)},\ldots,m_t^{(M)}\}$ contains the persistent
memories of all primitives and $h_t$ summarises recent realised behaviour. The
deployed actor receives a fixed-dimensional observation
\begin{equation}
o_t=[\hat z(G),E_\theta(\theta_t),\operatorname{onehot}(i_t),\rho_t,
E_m(\mathcal M_t),u_t],
\label{eq:actor-observation}
\end{equation}
where $u_t$ contains inexpensive trajectory features such as recent strategy
movement, update magnitude/alignment, reversal or oscillation statistics, and
time since the last switch. Exact exploitability and counterfactual inactive-
solver proposals are not exposed to the deployed actor.

The policy is hierarchical but simple. A gate samples
$g_t\sim\pi_{\mathrm{gate},\omega_g}(\cdot\mid o_t)$. If it chooses
\textsc{Stay}, the active primitive continues. If it chooses \textsc{Switch}, a
router samples $j_t\sim\pi_{\mathrm{route},\omega_r}(\cdot\mid o_t,i_t)$ from
$\mathcal I\setminus\{i_t\}$. The current primitive is masked from the router so
that staying remains an explicit decision. Each episode starts from a robust
reference primitive selected on development data (Optimistic in our experiments),
but the first control decision is taken at $C=0$, so the controller can depart
immediately when the game and warm start favour another mechanism.

Switching changes control, not solver identity. Primitive memories persist while
inactive, and they may be updated from quantities already generated by the
realised trajectory when doing so requires no extra operator evaluation. We do
not run uncharged shadow trajectories. Thus returning to an optimistic,
averaging or fictitious-play primitive resumes a meaningful state rather than
silently restarting it.

\subsection{Learning the controller}
\label{sec:controller-learning}
The actor should not need to evaluate every inactive primitive before making a
deployment-time decision. During training, however, such counterfactuals can
make value estimation easier. For primitive $j$, let
\begin{equation}
\Delta_t^{(j)}=T_j(G,\theta_t,m_t^{(j)})-\theta_t,
\label{eq:proposal}
\end{equation}
where $T_j$ denotes the strategy component of $\mathcal T_j$ for one update, and let
$q_t^{(j)}=\widetilde E_\Delta(\Delta_t^{(j)})$. The privileged critic receives
\begin{equation}
X_t^+=[o_t,q_t^{(1)},\ldots,q_t^{(M)},c_t^{(1)},\ldots,c_t^{(M)}],
\label{eq:privileged-state}
\end{equation}
while the gate and router see only $o_t$. We use a state-value critic
$V_\nu(X_t^+)$; the counterfactual proposals are discarded at deployment.

The controller is optimised against the reward in Eq.~\eqref{eq:reward}. The
target return is undiscounted, $G_t=\sum_{k=t}^{T-1}r_k$ with $\gamma=1$, while
GAE may use a separate bias--variance parameter $\lambda_{\mathrm{GAE}}$. The
router loss is evaluated only on timesteps at which the gate chose
\textsc{Switch}. We initialise the policy by behavioural cloning from a rolling
oracle that scores candidate solver choices by downstream AUC, then fine-tune
with PPO. This separates discovery of a useful switching structure from the
subsequent policy optimisation.

\subsection{Training sequence}
\label{sec:training}
Training proceeds in three stages. We first freeze the behavioural recogniser
and train the action-equivariant base initialiser. We then train the dynamic
controller from these starts, jointly across $3\times3$, $4\times4$ and
$5\times5$ games. Finally, the controller is frozen and the value-guided
procedure in Section~\ref{sec:value-guided-init} refines only the initialiser.
The final model pairs this refined initialiser with the unchanged controller.
This one-way refinement avoids the instability we observed when repeatedly
alternating the two parameter blocks while still making the starting point aware
of the downstream computational policy. Switching cost and all other
hyperparameters are selected on development games and frozen before
evaluation. Pseudocode is provided in Appendix~H.

\section{Optimality of Budgeted Solver Switching}
\label{sec:theory}

The switching problem admits a direct dynamic-programming interpretation as a
finite-horizon intervention problem, following the intervention-control
construction of LICRA~\cite{mguni2023licra}. Our objective is undiscounted: every solver
segment consumes a positive number of operator calls and the episode terminates
when the compute budget is exhausted. We work directly with the
remaining budget rather than introducing a discount solely to obtain a
contraction. Full proofs are given in
Appendix~A.

Let \(b_t=C_{\max}-C_t\) denote the remaining operator-call budget. At each
control point, an action determines which primitive governs the next control
segment; the primitive is then executed for the complete updates that fit before
the next control point or the end of the remaining budget. For an augmented
state \(X\) whose active primitive is \(i\), let
\[
\mathcal A_b(X)
=
\{\textsc{Stay}\}
\cup
\{\operatorname{switch}(j):j\neq i\},
\]
restricted to actions whose resulting segment is feasible within budget \(b\).
Let \(r(X,a)\) denote the reward accumulated over that segment,
\(c(X,a)\in\mathbb N_{+}\) its operator-call cost, and
\(P(\mathrm dX'\mid X,a)\) the induced transition kernel to the next control
state. For remaining budget \(b\), define
\begin{equation}
V_b^\star(X)
=
\sup_\pi
\mathbb E_\pi
\left[
\sum_{t\geq 0} r(X_t,a_t)
\,\middle|\,
X_0=X,\ b_0=b
\right],\; V_0^\star\equiv0
\label{eq:budget-value}
\end{equation}
where the sum terminates when no feasible segment remains.

\paragraph{Stay and switch values.}
For any admissible action \(a\), define
\begin{align}
Q_b^\star(X,a)={}r(X,a)\notag
+\int V_{b-c(X,a)}^\star(X')
P(\mathrm dX'\mid X,a).
\label{eq:budget-q}
\end{align}
Let \(a^{\mathrm{stay}}\) continue the current primitive and
\(a_j=\operatorname{switch}(j)\). Then
\(Q_b^{\mathrm{stay}}(X)=Q_b^\star(X,a^{\mathrm{stay}})\), while the best
intervention is
\begin{equation}
\mathcal M_bV^\star(X)
=
\max_{\substack{j\neq i\\c(X,a_j)\le b}}
Q_b^\star(X,a_j).
\label{eq:intervention-operator}
\end{equation}
Hence
\begin{equation}
V_b^\star(X)
=
\max\left\{
Q_b^{\mathrm{stay}}(X),
\mathcal M_bV^\star(X)
\right\}.
\label{eq:switch-bellman}
\end{equation}

\begin{theorem}[Finite-budget convergence and optimality]
\label{thm:finite-budget}
Suppose the primitive set is finite, one-step segment rewards are bounded and
measurable, every executed segment consumes a positive integer number of
operator calls, and the induced transition kernels are measurable. Then Eq.~\eqref{eq:switch-bellman} has a unique bounded measurable solution
for every \(b\in\{0,\ldots,C_{\max}\}\). Backward Bellman updates from
\(V_0^\star\equiv0\) recover these values after finitely many budget layers,
and any measurable selector attaining the Bellman maximum is an optimal
budget-feasible switching policy.
\end{theorem}

\paragraph{Proof strategy.}
The argument is a backward induction over the remaining operator-call budget.
At \(b=0\), the value is uniquely zero. At any \(b>0\), every feasible action
uses at least one call, so its continuation value depends only on value
functions at strictly smaller budgets, which have already been determined.
The finite action set then makes the Bellman maximum well defined and attained.
The Bellman principle shows that selecting a maximising action and subsequently
following optimal lower-budget decisions is optimal. Because the remaining
budget decreases strictly after every segment, this recursion terminates after
finitely many budget layers. Appendix~A gives the complete
boundedness, measurability, uniqueness and measurable-selection argument.

\begin{proposition}[Intervention condition]
\label{prop:obstacle}
At state \((X,b)\), switching is optimal whenever
\begin{equation}
\mathcal M_bV^\star(X)
>
Q_b^{\mathrm{stay}}(X),
\label{eq:obstacle-condition}
\end{equation}
and staying is optimal when the reverse inequality holds. Conditional on
switching, an optimal target primitive is any maximiser in
Eq.~\eqref{eq:intervention-operator}.
\end{proposition}

\paragraph{Proof strategy.}
Equation~\eqref{eq:switch-bellman} separates the optimal decision into two
branches: continuing the active primitive and intervening by transferring
control to the best alternative primitive. The sign of
\[
\mathcal M_bV^\star(X)-Q_b^{\mathrm{stay}}(X)
\]
therefore determines whether intervention or continuation is optimal, while the
maximiser in Eq.~\eqref{eq:intervention-operator} determines the optimal target
conditional on switching. The equality case and the corresponding set of
optimal actions are treated formally in Appendix~A.

The proposition gives the hierarchical gate/router architecture a direct control
interpretation. The gate approximates whether the value of intervention exceeds
the value of continuation, while the router approximates the maximising
primitive conditional on intervention. These results characterise the
underlying finite-budget switching problem; they do not establish global
convergence of the nonlinear PPO parameterisation used to approximate its
policy. We therefore compare the learned controller empirically with restricted
exact switching searches rather than claiming that PPO recovers the Bellman
optimum. Appendix~G.3 reports the full \(K\leq3\) oracle analysis.

\section{Experiments}
\label{sec:experiments}

We structure the evaluation around five questions. \textbf{Q1:} Does a strong
starting point make adaptive control unnecessary? \textbf{Q2:} Does the
controller need feedback from the realised trajectory, rather than only a game-
level solver choice? \textbf{Q3:} Do controller-aware initialisation and dynamic
control add up? \textbf{Q4:} Do the gains persist beyond the training horizon?
\textbf{Q5:} Does adaptive control transfer to larger and held-out game sizes?
Q1 separates start quality from adaptive refinement; Q2 isolates the value
of trajectory feedback; Q3 tests whether improved initialisation and dynamic
control are complementary; Q4 tests transfer beyond the training horizon; and
Q5 tests transfer across game sizes.
Appendix G provides the corresponding initialisation,
ablation, oracle, representation and compute analyses, with reproducibility
details in Appendix I.

\subsection{Setup}
\label{sec:exp-setup}

\paragraph{Games and evaluation panels.}
We use two-player general-sum normal-form games of sizes $3\!\times\!3$,
$4\!\times\!4$ and $5\!\times\!5$, normalised as in
Section~\ref{sec:problem}. TEST3 (2,048 games per size) was consulted during
development and is used for diagnostic comparisons and
illustrative trajectories. \rev{Two further panels were generated after the models they evaluate were frozen: an ablation panel (512 games per size; Table~\ref{tab:feedback-ablation}) and an evaluation panel (1,024 games per size; Tables~\ref{tab:refined-main} and~\ref{tab:budget-refined}).} Results are never pooled across
panels. All reported results average over three seeds.

\paragraph{Solvers and compute.}
The controller (\emph{Full}) chooses among seven stateful primitives
(Optimistic, extragradient, fictitious play, GDA, Mirror, a damped projected
update labelled Proximal in the implementation, and Anchoring) every $\Delta=10$ operator calls, with memories preserved across
switches. Extragradient costs two calls per update; the others cost one. Full
is trained jointly on all three sizes with budget $C=200$. \rev{Appendix~K compares smaller primitive libraries for the multi-size controller of Section~\ref{sec:scale-transfer}.}

\paragraph{Initialisers.}
\emph{Original Atlas} is the base warm-start head on the frozen behavioural
encoder; \emph{Refined Atlas} is the same head after the value-guided update of
Section~\ref{sec:value-guided-init}. \rev{Refinement uses $\beta=0.1$, learning rate $3\times10^{-4}$ and at most 24 epochs, selected on development games (Appendix~D).} As an external comparator we use
NfgTransformer~\cite{liu2024nfgtransformer}; \emph{NfgT} is trained on the same games as the Atlas head. Our reference baseline \emph{M1} is Original Atlas followed by fixed Optimistic.

\paragraph{Metrics and statistics.}
Our primary metric is the current-iterate Nash-gap area
\begin{equation}
\mathrm{AUC}_C(G,\pi)=\frac{1}{C}\int_0^C
\Gap_G(\theta_\pi(c))\,\mathrm dc,
\label{eq:exp-auc}
\end{equation}
which rewards reaching a low gap early and staying there. We mainly report $R_C(A)=\overline{\mathrm{AUC}}_C(A)/\overline{\mathrm{AUC}}_C(\mathrm{M1})$ (lower is better). Ablations report the relative reduction $1-\overline{\mathrm{AUC}}_C(\mathrm{Full})/
\overline{\mathrm{AUC}}_C(B)$ against comparator $B$. The \emph{hardest 5\%}
are the games with the highest M1 AUC within each size. Intervals are 95\% paired root-bootstrap intervals after averaging the three
training seeds per root. Table~\ref{tab:feedback-ablation} uses 2,000 resamples;
other main/value-guided comparisons use 10,000. Holm correction covers the six
Full-versus-One-later-switch/PPO-without-BC tests across game sizes; other
intervals are pointwise unless stated otherwise.

\subsection{Does a strong start make adaptive control unnecessary?}
\label{sec:exp-complementarity}

No. A better start and adaptive control help on different games. On the TEST3 analysis panel, NfgT begins far closer to equilibrium than the Original Atlas start, with an initial gap three to nine times smaller. Yet averaged over all games, this much better start with fixed Optimistic gives about the same benefit as adaptive control from the original start.

\begin{figure*}[t]
\centering
\includegraphics[width=0.79\textwidth]{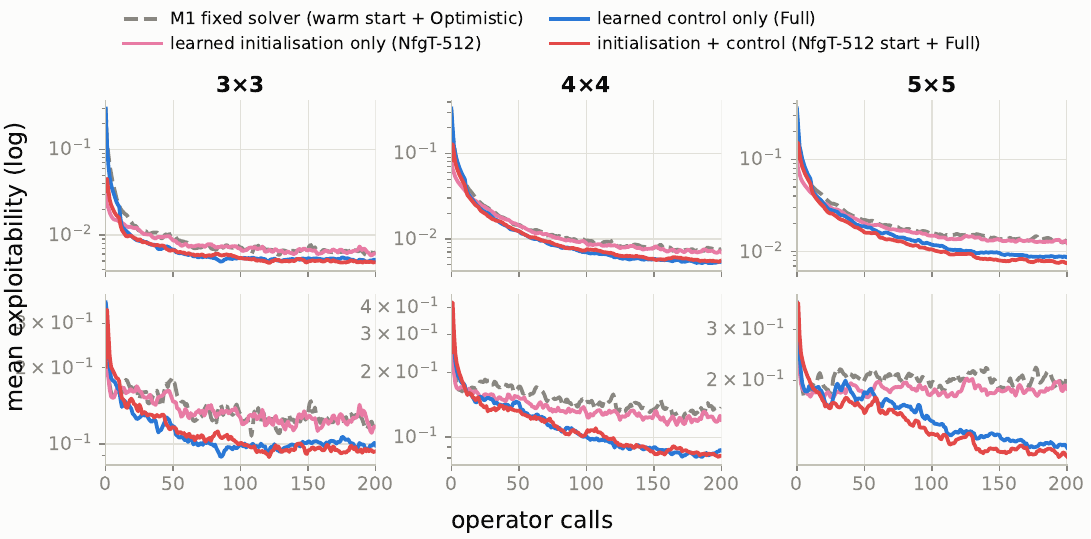}
\caption{\textbf{Start quality and adaptive control help on different games}
($C=200$, TEST3 analysis panel). Mean current-iterate Nash gap against operator calls. Top: all
games. Bottom: hardest 5\% by M1 AUC, selected within each size. A strong start
(NfgT + Optimistic) dominates on typical games; on hard games, adaptive
control (Full) makes the larger difference.}
\label{fig:aligned-complementarity}
\end{figure*}

Figure~\ref{fig:aligned-complementarity} shows that the two achieve this in very different places. On typical games, the strong start wins clearly. On the hardest 5\%, the order
reverses: Full reaches $R_{200}\approx0.75$--$0.80$, while the strong start with fixed Optimistic stays at $0.91$--$0.98$. A stronger start helps on typical games, but less where the fixed solver stalls or oscillates.

The two effects are complementary. Giving the NfgT start to the frozen Full controller improves on Original Atlas + Full overall, while on the hardest
5\% the two are indistinguishable (Appendix~G.1). This points to trajectory feedback, rather than start quality alone, as the important factor
on these difficult games.

\subsection{Does the controller need feedback from the trajectory?}
\label{sec:exp-feedback}

Yes. Table~\ref{tab:feedback-ablation} compares Full with controllers that receive progressively less information. \emph{Static selection} sees only the
game and picks one primitive for the whole solve; it is 12--15\% worse than Full at every size. An \emph{open-loop} controller also sees elapsed compute, so it can follow a fixed switching schedule, but it cannot see how the solve is going. It closes part of the gap, yet remains worse than Full on $4\!\times\!4$ and $5\!\times\!5$; on $3\!\times\!3$ the difference is not significant. On the hardest 5\%, its penalty is two to three times larger than on average. Knowing the game, and even knowing when to switch in general, is therefore not enough. The controller needs to see what is happening in this particular solve, and this matters most on the games
where control mattered most in Section~\ref{sec:exp-complementarity}.

\begin{table}[!t]
\centering
\caption{\textbf{Controller ablations on the ablation panel.} Relative AUC reduction (\%) of Original Atlas + Full over separately trained variants ($C=200$), with 95\% paired bootstrap intervals; positive values favour Full. Static selection and Open-loop remove trajectory feedback; One later switch uses Full's frozen $C=0$ opening decision and permits at most
one subsequent switch; PPO without BC removes behavioural-cloning initialisation.}
\label{tab:feedback-ablation}
\begingroup
\setlength{\tabcolsep}{3.2pt}
\footnotesize
\begin{tabular}{lccc}
\toprule
Comparator & $3\!\times\!3$ & $4\!\times\!4$ & $5\!\times\!5$\\
\midrule
Static selection      & 11.7 [6.3,17.6] & 14.6 [6.2,21.9] & 14.0 [6.9,20.8]\\
Open-loop (game+time) & 4.5 [-0.2,9.0]  & 9.7 [1.3,16.9]  & 7.7 [3.4,11.9]\\
One later switch            & 4.5 [-0.7,9.2]  & 10.8 [2.9,18.0] & 9.3 [4.5,14.0]\\
PPO without BC        & 10.2 [5.7,15.0] & 9.9 [1.1,18.1]  & 11.5 [5.8,16.7]\\
\bottomrule
\end{tabular}
\endgroup
\vspace{-10pt}
\end{table}

How many switches does good control need? Only a few. Exact searches over schedules with at most three switches show that most of the
attainable improvement comes from choosing the right primitive at the start, and that a single well-timed later switch captures most of what further
switching adds (Figure~\ref{fig:phase-dependent}a; Section~G.3). The learned controllers agree: One later switch is already strong, although Full still improves on it for $4\!\times\!4$ and $5\!\times\!5$. Initialising PPO by behavioural cloning from the rolling oracle also matters: without it, the controller is 10--12\% worse.

\begin{figure*}[t]
\centering
\includegraphics[width=0.79\textwidth]{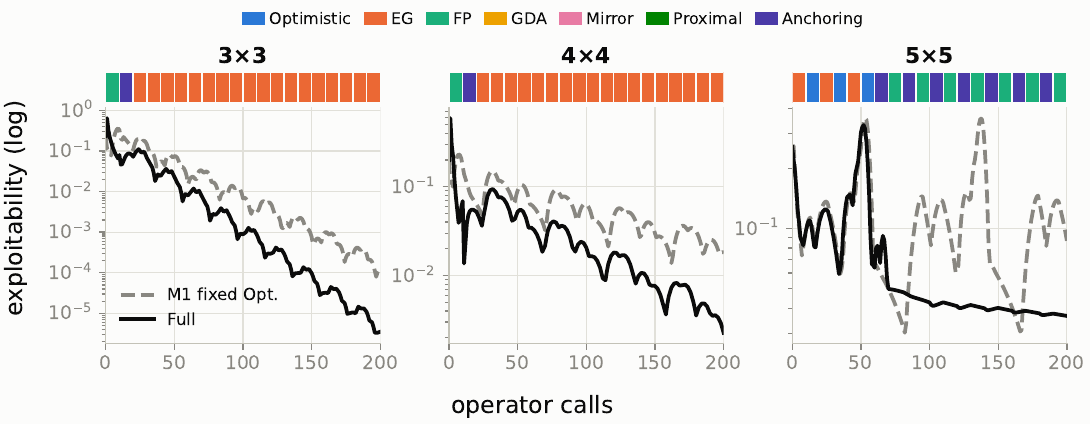}
\caption{\textbf{Phase-dependent control on hard games from the analysis panel} ($C=200$). Top strips show the primitive used by Full in each ten-call
window; lower panels compare its Nash gap with M1. Example games were selected
by a rule fixed before plotting. Left and centre: early decisions followed
by commitment. Right: continued adjustment until the solve stabilises.}
\label{fig:example-trajectories}
\end{figure*}

Figure~\ref{fig:example-trajectories} shows what these decisions look like on hard games from the TEST3 analysis panel. In the $3\!\times\!3$ and $4\!\times\!4$ examples, Full
tries a few mechanisms early and then commits to extragradient. In the $5\!\times\!5$ example, where fixed Optimistic keeps oscillating, Full keeps changing mechanism until the gap stabilises and then avoids the later spikes. These are single trajectories and do not justify every individual switch, but they make the phase dependence visible.

\subsection{Do better initialisation and adaptive control add up?}
\label{sec:exp-main}

Yes. Table~\ref{tab:refined-main} crosses the two starts (Original and Refined Atlas) with two downstream
solver policies (fixed Optimistic and Full). \rev{Each component improves on M1 on its own, and their combination gives the lowest AUC in Table~\ref{tab:refined-main} at every size, recovering 85--92\% of the sum of the two individual gains. From the Refined start, Full lowers AUC by a further 23.8\%, 7.2\% and 12.2\%, significantly so at $3\times3$ and $5\times5$. Refined Atlas + Full is also significantly better than NfgT + Full at every size; Appendix~G.1 compares against an NfgT trained on 24 times more games.} A better start therefore does not do the job that the controller does.

\begin{table*}[t]
\centering
\caption{\textbf{Main result on the evaluation panel} ($C=200$, \rev{1,024 games per size}, three seeds). Initial gap is measured before any solver call;
$R_{200}$ is AUC relative to M1 on the same games (lower is better). The top
block is a $2\times2$ design over start (Original/Refined) and downstream solver policy
(fixed Optimistic/Full).}
\label{tab:refined-main}
\begingroup
\setlength{\tabcolsep}{4.1pt}
\footnotesize
\begin{tabular}{lrrrrrr}
\toprule
&\multicolumn{3}{c}{Initial Nash gap}&\multicolumn{3}{c}{Normalised AUC $R_{200}$}\\
Method & $3\!\times\!3$ & $4\!\times\!4$ & $5\!\times\!5$ & $3\!\times\!3$ & $4\!\times\!4$ & $5\!\times\!5$\\
\midrule
Original Atlas + Optimistic (M1) & \rev{.297} & \rev{.335} & \rev{.352} & \rev{1.000} & \rev{1.000} & \rev{1.000}\\
Original Atlas + Full & \rev{.297} & \rev{.335} & \rev{.352} & \rev{.772} & \rev{.895} & \rev{.878}\\
Refined Atlas + Optimistic & \rev{.026} & \rev{.060} & \rev{.075} & \rev{.737} & \rev{.792} & \rev{.768}\\
\textbf{Refined Atlas + Full (ours)} & \textbf{\rev{.026}} & \textbf{\rev{.060}} & \textbf{\rev{.075}} & \textbf{\rev{.562}} & \textbf{\rev{.735}} & \textbf{\rev{.674}}\\
\midrule
NfgT + Optimistic & \rev{.033} & \rev{.082} & \rev{.098} & \rev{.779} & \rev{.844} & \rev{.862}\\
NfgT + Full & \rev{.033} & \rev{.082} & \rev{.098} & \rev{.638} & \rev{.781} & \rev{.784}\\
\bottomrule
\end{tabular}
\endgroup
\end{table*}

\rev{The gain from refinement is concentrated at the start of the solve. Relative to Original Atlas + Full, most of the improvement occurs in the first 20 calls, with a lower gap on average over the remaining calls. Value guidance improves the state from which the controller starts.}

\subsection{Do the gains persist beyond the training horizon?}
\label{sec:exp-budget}

Yes. We evaluate the same frozen models, trained at $C=200$, at $C=400$ and $C=1000$ without retraining (Table~\ref{tab:budget-refined}). \rev{The full pipeline retains its AUC advantage over M1 at both longer horizons, and the refined start continues to improve performance relative to Original Atlas + Full. The advantage of the Refined start narrows as the horizon grows, consistent with refinement acting mainly in the first calls of the solve.}

\begin{table}[!t]
\centering
\caption{\textbf{Budget transfer on the evaluation panel.} Normalised
current-iterate AUC of the same frozen models trained at $C=200$ and evaluated
without retraining at the stated horizons; lower is better. Bold marks our method.}
\label{tab:budget-refined}
\begingroup
\setlength{\tabcolsep}{3.0pt}
\footnotesize
\begin{tabular}{llccc}
\toprule
Budget & Method & $3\!\times\!3$ & $4\!\times\!4$ & $5\!\times\!5$\\
\midrule
$400$ & Original Atlas + Full & \rev{.777} & \rev{.867} & \rev{.815}\\
 & \textbf{Refined Atlas + Full} & \textbf{\rev{.652}} & \textbf{\rev{.757}} & \textbf{\rev{.667}}\\
 & NfgT + Full & \rev{.701} & \rev{.792} & \rev{.761}\\
\midrule
$1000$ & Original Atlas + Full & \rev{.846} & \rev{.772} & \rev{.778}\\
 & \textbf{Refined Atlas + Full} & \textbf{\rev{.776}} & \textbf{\rev{.722}} & \textbf{\rev{.649}}\\
 & NfgT + Full & \rev{.756} & \rev{.749} & \rev{.724}\\
\bottomrule
\end{tabular}
\endgroup
\end{table}

\subsection{\rev{Does adaptive control transfer to larger and held-out sizes?}}
\label{sec:scale-transfer}
\rev{To separate size transfer from learned initialisation, we train a single controller, Full-L (three seeds), on games with 3, 4, 5, 10, 20, 50 and 100 actions and evaluate it from uniform starts. Full-L uses the gate/router and seven-primitive library of Full, with size-normalised observation statistics. Sizes 15, 30 and 75 are held out from controller training and lie within the training range. Baselines are tuned on training and development games of each evaluation size, and all models are frozen before the evaluation games are generated.}

\rev{At $C=200$, Full-L is significantly better in 20 of the 30 comparisons with a fixed primitive selected on training games, kNN static selection and stall-based switching, and worse in two (Holm correction; Appendix~J). It improves on all three comparators at the held-out sizes 15 and 30; gains over kNN are approximately 5--10\% at 10--30 actions. Its mean AUC is also lower than that of every individual fixed primitive at every size, by 2--13\%. Trajectory-conditioned switching therefore transfers to unseen sizes without a learned start.}

\section{Discussion}
\label{sec:discussion}

The experiments distinguish two roles. Initialisation transfers information
about where equilibrium search should begin, while dynamic control determines
how solver mechanisms should be used once the trajectory unfolds. Value-guided
refinement improves the starting state without removing the benefit of later control. The static-selection and One-later-switch ablations further show that
game-level solver choice is insufficient, although a small number of
well-timed interventions captures much of the available gain.

Our experiments are limited to a fixed primitive library and finite-horizon
normal-form games. The frozen behavioural representation provides game-level
context rather than predicting exact switch timing, and controller inference
adds computational overhead beyond the operator calls used for solver
accounting. At larger sizes the controller optimises trajectory AUC and reaches stringent terminal tolerances less often than extragradient, and with in-memory payoffs its decision and representation costs are repaid only in the regimes identified in Appendix~J. Broader game families, including many-player games whose equilibria can be
approximated in the limit of infinitely many agents~\cite{mguni2018manyagents},
larger primitive libraries and direct
training at longer horizons remain for future work.
\section{Conclusion}
\label{sec:conclusion}

Repeated game solving creates two opportunities for amortisation: learning a better place to start and learning how to adapt the solver as optimisation unfolds. Our framework combines a controller-aware warm start with trajectory-dependent switching among persistent solver mechanisms. On held-out games, the complete pipeline reduces normalised exploitability AUC by \rev{26.5--43.8\%} relative to the original warm start with fixed Optimistic, while dynamic control continues to contribute after the start is refined and improves over static per-game solver selection. These results support treating solver choice as a learned, state-dependent
part of equilibrium computation rather than fixing it before the solve begins.

\bibliographystyle{ACM-Reference-Format}
\bibliography{sample}

\appendix
\addtocontents{toc}{\protect\setcounter{tocdepth}{2}}
\onecolumn

\setcounter{tocdepth}{2}
\tableofcontents
\clearpage
\section{Proofs of the theoretical results}
\label{app:proofs}

This appendix gives the full proofs of the dynamic-programming results stated in
the main paper. The proof architecture follows the intervention-control
constructions used by LICRA~\cite{mguni2023licra} and the later multi-mode
intervention formulation of MINT~\cite{mguni2025mint}: the value is decomposed
into a continuation branch and one or more intervention branches, and the
optimal decision is characterised by their Bellman comparison. There is,
however, one important difference. Those analyses establish convergence in a
discounted infinite-horizon setting through a contraction argument. Our
equilibrium-computation objective is undiscounted but has a finite
operator-call budget. We therefore use backward induction on the remaining
budget rather than introducing an artificial discount factor.

\subsection{Preliminaries}

Let \(X\) denote the augmented solver state, including the game, current
strategy, active primitive, persistent primitive memories and the trajectory
summary required by the transition model. Let \(b\in\{0,\ldots,C_{\max}\}\)
denote the remaining operator-call budget. If the active primitive is \(i\), the
budget-feasible action set is
\[
\mathcal A_b(X)
=
\left\{\textsc{Stay}\right\}
\cup
\left\{\operatorname{switch}(j):j\neq i\right\},
\]
with actions whose executed primitive costs more than \(b\) removed. We write
\(c(X,a)\in\mathbb N_+\) for the operator-call cost of an admissible action,
\(r(X,a)\) for its one-step reward, and
\(P(\mathrm dX'\mid X,a)\) for the induced transition kernel. If no update is
budget feasible, the process terminates.

In the experiments, a controller decision selects the primitive for a window
of ten operator calls. The results below apply with each action interpreted
as an entire window: $c(X,a)$ counts all calls executed in that window
(ten, or fewer in a final partial window). The full state $X$ contains
all primitive memories and trajectory statistics required for the Markov
transition. The deployed actor sees only $o_t=o(X_t)$; the theorem concerns
the full-state control problem, not optimality of the learned actor.

For full measure-theoretic precision, the theorem in the main paper is
understood with the standard additional requirement that \(r(\cdot,a)\) and
\(c(\cdot,a)\) are measurable for every action \(a\). This holds automatically
for the implemented finite primitive library and deterministic operator-call
costs. We assume throughout that there is \(R_{\max}<\infty\) such that
\[
|r(X,a)|\le R_{\max}
\]
for every admissible state--action pair.

\subsection{Proof of Theorem~1: finite-budget convergence and optimality}

We restate the result using the notation above.

\begin{theorem}[Finite-budget convergence and optimality]
\label{thm:supp-finite-budget}
Assume that the primitive set is finite, one-step rewards are bounded and
measurable, every executed primitive consumes a measurable positive integer
number of operator calls, and the induced transition kernels are measurable.
Then the budgeted Bellman recursion
\[
V_b^\star(X)
=
\max_{a\in\mathcal A_b(X)}
\left\{
r(X,a)
+
\int
V_{b-c(X,a)}^\star(X')
P(\mathrm dX'\mid X,a)
\right\},
\]
with \(V_0^\star\equiv0\) and terminal value zero whenever no update is
admissible, has a unique bounded measurable solution for every
\(b\in\{0,\ldots,C_{\max}\}\). Backward Bellman updates from
\(V_0^\star\) recover these value functions after finitely many budget layers,
and a measurable selector attaining the Bellman maximum is an optimal
budget-feasible switching policy.
\end{theorem}

\begin{proof}
The proof is by induction on the remaining operator-call budget \(b\).

\paragraph{Base case.}
For \(b=0\), no primitive can be executed because every admissible primitive has
strictly positive integer cost. The episode therefore terminates and
\[
V_0^\star(X)=0
\]
for every state \(X\). This solution is bounded, measurable and unique.

\paragraph{Induction hypothesis.}
Fix \(b\ge1\) and suppose that for every \(b'<b\), the function
\(V_{b'}^\star\) has already been uniquely defined, is bounded and measurable,
and equals the optimal expected return achievable with remaining budget \(b'\).

\paragraph{One-step action values are well defined.}
For an admissible action \(a\in\mathcal A_b(X)\), define
\[
Q_b^\star(X,a)
=
r(X,a)
+
\int
V_{b-c(X,a)}^\star(X')
P(\mathrm dX'\mid X,a).
\]
Since \(c(X,a)\ge1\), the continuation budget satisfies
\(b-c(X,a)<b\). The continuation value appearing on the right-hand side is
therefore already uniquely defined by the induction hypothesis. Measurability
of the transition kernel and of \(V_{b-c(X,a)}^\star\) implies measurability of
the integral, and hence of \(Q_b^\star(\cdot,a)\).

Boundedness also follows immediately. If
\(\|V_{b'}^\star\|_\infty\le b'R_{\max}\) for every \(b'<b\), then
\[
|Q_b^\star(X,a)|
\le
R_{\max}
+
\|V_{b-c(X,a)}^\star\|_\infty
\le
R_{\max}+(b-1)R_{\max}
=
bR_{\max}.
\]
Thus every feasible action value is bounded.

\paragraph{Bellman value and measurable maximiser.}
The primitive library is finite, so the maximum
\[
\widehat V_b(X)
=
\max_{a\in\mathcal A_b(X)}Q_b^\star(X,a)
\]
is attained whenever an action is feasible. A finite maximum of measurable
functions is measurable, and the preceding bound gives
\(\|\widehat V_b\|_\infty\le bR_{\max}\).
Because the candidate \(Q_b^\star\) values are uniquely determined by lower
budget layers, \(\widehat V_b\) is itself uniquely determined.

A measurable maximising selector exists in this finite-action setting. Fix an
ordering of the controller actions and choose the first action attaining the
maximum. The set on which a particular action is the first maximiser is an
intersection of measurable pairwise comparison sets, hence is measurable.
Denote this selector by \(a_b^\star(X)\).

\paragraph{Dynamic-programming upper bound.}
Consider any budget-feasible policy \(\pi\) starting from \((X,b)\), and let its
first action be \(a\). After that action, the remaining budget is
\(b-c(X,a)<b\). By the induction hypothesis, no continuation policy can obtain
more expected return than \(V_{b-c(X,a)}^\star\). Consequently,
\[
J_b^\pi(X)
\le
r(X,a)
+
\int
V_{b-c(X,a)}^\star(X')
P(\mathrm dX'\mid X,a)
=
Q_b^\star(X,a)
\le
\widehat V_b(X).
\]
Taking the supremum over all feasible \(\pi\) yields
\(V_b^\star(X)\le\widehat V_b(X)\).

\paragraph{Achievability.}
Now choose \(a_b^\star(X)\) at the first decision. For every realised successor
state \(X'\), continue with an optimal selector for the smaller budget
\(b-c(X,a_b^\star(X))\), whose existence is guaranteed by the induction
hypothesis. This policy attains
\[
Q_b^\star(X,a_b^\star(X))
=
\widehat V_b(X),
\]
so \(V_b^\star(X)\ge\widehat V_b(X)\). Combining the two inequalities gives
\[
V_b^\star(X)=\widehat V_b(X).
\]
Thus the Bellman equation holds at budget \(b\), its bounded measurable solution
is unique, and the maximising selector followed by optimal lower-budget
selectors is optimal.

By induction, the claim holds for every
\(b=0,\ldots,C_{\max}\).

\paragraph{Finite recovery by Bellman updates.}
Unlike the discounted intervention models in
\cite{mguni2023licra,mguni2025mint}, no asymptotic fixed-point iteration is
required here. Every Bellman update at budget \(b\) depends only on already
computed layers with strictly smaller remaining budget. Hence
\(V_1^\star,V_2^\star,\ldots,V_{C_{\max}}^\star\) are recovered successively in
at most \(C_{\max}\) budget layers. This proves finite convergence and
optimality.
\end{proof}

\begin{corollary}[Budget feasibility and finite termination]
\label{cor:supp-budget}
Under the assumptions of Theorem~\ref{thm:supp-finite-budget}, any policy whose
actions are restricted to \(\mathcal A_b(X)\) satisfies the operator-call budget
almost surely and terminates after at most \(C_{\max}\) executed primitive
updates.
\end{corollary}

\begin{proof}
At each executed update,
\[
b_{t+1}=b_t-c(X_t,a_t),
\qquad
1\le c(X_t,a_t)\le b_t.
\]
Thus \(b_t\) remains non-negative and decreases by at least one whenever an
update is executed. Starting from \(b_0=C_{\max}\), no more than
\(C_{\max}\) updates can occur. Since actions with cost exceeding the remaining
budget are removed from the feasible set, the budget cannot be violated.
\end{proof}

\subsection{Proof of Proposition~1: the intervention condition}

The stay/switch decomposition is the direct analogue of the obstacle condition
in impulse-control formulations~\cite{mguni2023licra}. The difference is that
the intervention here changes the active equilibrium-solving primitive rather
than triggering an external impulse.

Recall
\[
Q_b^{\mathrm{stay}}(X)
=
Q_b^\star(X,a^{\mathrm{stay}})
\]
and
\[
\mathcal M_bV^\star(X)
=
\max_{\substack{j\neq i\\c(X,a_j)\le b}}
Q_b^\star(X,a_j),
\]
where \(i\) is the active primitive. If no switch is budget feasible, we set \(\mathcal M_bV^\star(X)=-\infty\);
if staying is not budget feasible, we set \(Q_b^{\mathrm{stay}}(X)=-\infty\). The Bellman equation is
\[
V_b^\star(X)
=
\max\left\{
Q_b^{\mathrm{stay}}(X),
\mathcal M_bV^\star(X)
\right\}.
\]

\begin{proposition}[Intervention condition]
\label{prop:supp-obstacle}
At state \((X,b)\):
\begin{enumerate}[label=(\roman*),leftmargin=*]
    \item if
    \(\mathcal M_bV^\star(X)>Q_b^{\mathrm{stay}}(X)\),
    every optimal first decision switches;
    \item if
    \(\mathcal M_bV^\star(X)<Q_b^{\mathrm{stay}}(X)\),
    every optimal first decision stays;
    \item if the two quantities are equal and finite, staying and every
    maximising switch are both optimal;
    \item conditional on switching, any
    \[
    j^\star
    \in
    \arg\max_{\substack{j\neq i\\c(X,a_j)\le b}}
    Q_b^\star(X,a_j)
    \]
    is an optimal target primitive.
\end{enumerate}
\end{proposition}

\begin{proof}
By Theorem~\ref{thm:supp-finite-budget}, the optimal value at \((X,b)\) is the
maximum of exactly two branches: continuation with the current primitive and
the best feasible switch.

If
\(\mathcal M_bV^\star(X)>Q_b^{\mathrm{stay}}(X)\),
then
\[
V_b^\star(X)=\mathcal M_bV^\star(X)
\]
and the stay branch is strictly suboptimal. Hence any optimal first action must
belong to the switching branch. Conversely, if
\(\mathcal M_bV^\star(X)<Q_b^{\mathrm{stay}}(X)\), then
\[
V_b^\star(X)=Q_b^{\mathrm{stay}}(X),
\]
so every switching action is strictly dominated by staying.

If equality holds, the two branches attain the same Bellman value, so either
branch is optimal, provided the switching maximum is attained by a feasible
primitive. Finally, once switching is selected, the definition of
\(\mathcal M_b\) shows that a target primitive is optimal if and only if it
attains the displayed maximum. This proves all four statements.
\end{proof}

\paragraph{Relation to the learned gate and router.}
The proposition gives the controller architecture a direct dynamic-programming
interpretation. The gate approximates the sign of the intervention advantage
\[
A_b^{\mathrm{sw}}(X)
=
\mathcal M_bV^\star(X)-Q_b^{\mathrm{stay}}(X),
\]
while, conditional on a positive switching decision, the router approximates
the maximising primitive in the intervention operator. As in LICRA's obstacle
condition and MINT's multi-mode Bellman maximum, the structural claim is about
the optimal control problem itself; it does not imply global convergence of the
nonlinear PPO parameterisation used to approximate the policy in the
experiments~\cite{mguni2023licra,mguni2025mint}.

\section{Game corpus, roots and evaluation panels}
\label{app:data}

\paragraph{Games and payoff scale.}
We consider two-player games of sizes $3\times3$, $4\times4$ and $5\times5$.
The controller data are generated from a hybrid distribution: even-numbered
seeds generate independent standard-Gaussian payoffs, while odd-numbered seeds
combine ten structured families with weights sampled from
$\mathrm{Dirichlet}(1,\ldots,1)$. The families are potential, zero-sum,
cyclic, symmetric, anti-coordination, general-sum,
degenerate-coordination, boundary-support, interior-support and
monotonicity-controlled games. The structured parameters are: potential dummy
scale $0.5$; noise $0.05$ for anti-coordination, boundary-support and
interior-support; noise $0.02$ for degenerate-coordination; boundary-support
margin $1$; and monotonicity parameter $\alpha=0.5$.
Each root generates one game per size. Acceptance is determined only from its
$3\times3$ instance, requiring
$a_{\rm mono}\in[-1.2782804413,-0.6841871142]$ and
$z_{\rm pot},z_{\rm harm},z_{\rm zs},z_{\rm sym}\in[0.1,0.9]$.
Acceptance need not preserve an equal proportion of the two generator branches.
The Atlas pretraining corpus instead uses structured coverage sampling, interpolation
and perturbation augmentation.
For each game, the two payoff matrices are centred separately and divided by
one common scale:
\[
 A'=\frac{A-\bar A}{s},\quad B'=\frac{B-\bar B}{s},\qquad
 s=\max\{\|A-\bar A\|_\infty,\|B-\bar B\|_\infty,10^{-8}\}.
\]
Thus the entries lie in $[-1,1]$. Solver evaluations use these cached payoffs
without further dataset-level rescaling. Encoder-internal transformations do not
change the payoff scale used for reported Nash gaps.

\paragraph{Independent roots.}
For the Atlas pretraining data, a root is traced recursively through the parent-game
relationship. Derived games can be formed by Gaussian payoff perturbations
followed by renormalisation; observed positive perturbation magnitudes are
$0.005,0.01,0.02,0.03,0.05,0.08,0.12,0.16$.
For the multi-size controller corpus, one generator draw is a root shared
by its three size-specific games. Hence 4,096 games per size represent 4,096
shared roots. Splits and bootstrap resampling operate at root level.

\begin{table}[t]
\centering
\caption{Data pools and evaluation panels. Counts refer to game instances unless
roots are stated explicitly.}
\label{tab:supp-panels}
\footnotesize
\begin{tabular}{@{}lp{0.25\textwidth}p{0.43\textwidth}@{}}
\toprule
Data & Size & Role \\
\midrule
Encoder pretraining & 20,000 $3\times3$ (1,068 roots) & Frozen game representation \\
Initialiser TRAIN & 512 $3\times3$ (313 roots) & Original initialiser head; matched transformer \\
Initialiser DEV & 128 $3\times3$ (106 roots) & Checkpoint selection \\
Controller TRAIN & 4,096 per size (12,288 total; 4,096 roots) & BC and PPO \\
Controller DEV & 512 per size (1,536 total; 512 roots) & Controller checkpoint selection \\
Refinement TRAIN/DEV & Same controller TRAIN/DEV & Surrogate fitting and warm-start refinement \\
Second development panel (DEV2) & 2,048 per size (6,144 total; 2,048 roots) & Development diagnostics and probes \\
TEST3 & 2,048 per size & Development-consulted analysis panel \\
Ablation panel & 512 roots per size & Controller ablations (main Table 1); generated after the controllers were frozen \\
\rev{Evaluation panel} & \rev{1,024 roots per size} & \rev{Main Tables 2 and 3; generated after all models were frozen} \\
Multi-size evaluation panel & 512 roots per size at 3, 4, 5, 10, 15, 20, 30 and 75 actions; 1,024 at 50 and 100 & Size transfer and portfolio ablation (Appendices~\ref{app:scale-cost} and~\ref{app:portfolio}); generated after all models were frozen \\
Timing subset & First 256 games per size of the multi-size evaluation panel at 20, 50 and 100 actions & Elapsed-time comparisons (Appendix~\ref{app:scale-cost}) \\
\bottomrule
\end{tabular}
\end{table}

The training stages have different objectives but not disjoint data: the
controller and refinement reuse the same multi-size TRAIN/DEV pools. Of 512
initialiser TRAIN instances, 408 also appear in encoder TRAIN, and all 313
roots overlap; all 128 initialiser DEV instances appear in encoder DEV (106
roots). Encoder TRAIN shares no game identities or roots with initialiser DEV.
Refinement produces nine rollouts per training game and seed (110,592 TRAIN
and 13,824 DEV rollouts), not additional independent games.

\section{Solver primitives and operator calls}
\label{app:primitives}

\paragraph{Operator-call accounting.}
At strategy $\theta=(x,y)$ one operator call evaluates
$g(\theta)=(Ay,B^{\top}x)$ on the two cached payoff matrices.
Extragradient requires two calls per update; other primitives require one.
Reading payoffs as encoder input is permitted: the operator-call budget counts
solver evaluations, not black-box access to the payoff matrices. Offline gap
measurement for AUC is not included as deployable actor input; explicit gap
certification is charged one call under the certification protocol.

\paragraph{Updates and shared memory.}
Let $P$ be the product-simplex Euclidean projection, $p$ the gradient at the
start of the last executed update, $h$ the running anchor, and $k$ the number
of executed updates so far. The seven implemented operators are:
\begin{table}[t]
\centering
\caption{Primitive update rules and size-dependent step sizes. Here
$g=g(\theta)$, $\mathrm{BR}$ is the best-response profile, and $P$ projects
onto both simplices.}
\label{tab:supp-primitives}
\footnotesize
\setlength{\tabcolsep}{3pt}
\begin{tabular}{@{}lp{0.49\textwidth}c@{}}
\toprule
Primitive & Update & $\eta$ for $3/4/5$ actions \\
\midrule
Optimistic & $P[\theta+\eta(2g-p)]$ & $0.5/0.5/0.5$ \\
Extragradient & $v=P(\theta+\eta g)$, $\theta^+=P[\theta+\eta g(v)]$ & $1/1/1$ \\
Fictitious play & $P[\theta+(\mathrm{BR}(\theta)-\theta)/(k+1)]$ & -- \\
GDA & $P(\theta+\eta g)$ & $.75/.375/.375$ \\
Mirror & $s_a^+\propto s_a\exp(\eta g_a)$ for each player & $1.25/.625/.625$ \\
Damped projected update & $P[.5\theta+.5P(\theta+\eta g)]$ & $2.5/.625/.625$ \\
Anchoring & $P[\theta+\eta g+.25(h-\theta)]$ & $.75/.75/.75$ \\
\bottomrule
\end{tabular}
\end{table}
The sixth primitive is labelled Proximal elsewhere for compatibility with the
implementation; it does not solve an implicit proximal subproblem. Step-size
multipliers were selected from $\{0.125,0.25,0.5,1,2\}$ using mean fixed-solver
$C=200$ AUC on the first 1,024 TRAIN games per size at seed 0.
Initial memory is $p_0=0$ and uniform $h_0$. After every realised primitive
update, regardless of its identity, set $p\gets g(\theta_{\rm old})$; an
extragradient look-ahead gradient does not replace this value. With the already
available gradient, compute the Anchoring candidate $a_k$ and update
$h_{k+1}=(kh_k+a_k)/(k+1)$, even when Anchoring is inactive, with no new
operator call. Fictitious play uses the global executed-update count $k$.
Other primitives have no separate momentum or look-ahead caches. The current
strategy, iteration count and trajectory features follow the realised updates;
switching does not reset them.

\section{Initialiser and controller implementation}
\label{app:implementation}

The deployed method has two learned components: a controller-aware initialiser and
a dynamic solver controller. The behavioural recogniser is frozen. The final
training sequence is deliberately one-way: train the base initialiser, train the
controller from those starts, freeze the controller, and then refine only the
initialiser using the local value model described in the main paper. The deployed
controller is not retrained after this final value-guided update.

\rev{The frozen atlas is built from 20,000 $3\times3$ games evaluated under 31 configurations: five step sizes for each of six solver mechanisms and one fictitious-play configuration, each run for 200 calls. A payoff recogniser predicts the resulting 12-dimensional atlas coordinates and is frozen before controller training. The representation supplies game-level context (Appendix~\ref{app:representation}); the primitive-portfolio ablation appears in Appendix~\ref{app:portfolio}.}

\paragraph{Base initialiser.}
The original Atlas initialiser is trained using 512 $3\times3$ games (313 roots),
with checkpoint selection on 128 development games. Its objective uses the
mean gap over 200 \emph{updates} of fixed Optimistic, rather than the later
trapezoidal operator-call AUC:
\[
\mathcal L_{\rm base} = \mathbb E_G\!\left[
\frac{200^{-1}\sum_{t=1}^{200}\Gap(\theta_t)}
{\max\{\Gap(\theta_{\rm uniform}),10^{-3}\}}\right].
\]
The denominator is detached. Adam uses learning rate $10^{-3}$, batch size
64, 20 epochs (160 optimiser steps) and gradient clipping at 1. The DEV
checkpoint minimises $L_{200}$, including epoch 0. The action-wise head is
applied without parameter changes to larger games.

\paragraph{Value-guided initialiser refinement.}
The procedure trains the base initialiser, trains the Full
controller from those starts, freezes that controller and then fine-tunes only
the initialiser using a fitted downstream objective model. Write $\theta_{\psi_0}(G)$ for the base start. We form $K=8$
perturbations plus the original start, each by applying softmax to player-wise
perturbed logits
\[
\widetilde\ell=\ell+\sigma\,\operatorname{std}(\ell)\,\varepsilon,
\quad\varepsilon\sim\mathcal N(0,I),\qquad
\sigma\in\{0.25,0.5,1\},
\]
where population standard deviation is across actions and $\sigma$ is cycled
by game and perturbation index. Each start is evaluated with the frozen
controller. Let $s_{n,r}$ be the TRAIN mean AUC of Original Atlas + fixed
Optimistic for size $n$ and training seed $r$. The fitted target is
\[
J_{\pi_\omega}(G,\theta_0)=
\frac{\AUC_{C_{\max}}(G,\pi_\omega\mid\theta_0)}{s_{n,r}}
+0.01 N_{\rm sw}.
\]
Thus $0.01$ is the switching coefficient on \emph{normalised} AUC; on raw
AUC its equivalent is $\lambda_{\rm sw}=0.01s_{n,r}$.
The regressor receives $[\hat z(G),\mathrm{onehot}(n),E_\theta(G,\theta_0)]$
of dimension $12+3+30=45$, where game-dependent strategy features comprise
per-player mean/min/max/RMS of strategy probabilities, action payoffs and
per-action regret relative to the player's expected payoff, plus expected
payoffs, maximum unilateral improvements, initial gap and
$\log(\Gap(\theta_0)+10^{-4})$.
The regressor has layers $45\to128\to128\to1$ with Tanh hidden activations.
TRAIN-derived input/target normalisation is used; its fitting loss combines
standardised MSE and within-game pairwise ranking loss with equal weights.
Adam uses $10^{-3}$, batch size 256 games, up to 300 epochs and DEV patience
30. With both controller and regressor frozen, only the 9,985 parameters of
the action encoder/head are updated by minimising
\[
\mathcal L_{\rm VG}(\psi)=\mathbb E_G\!\left[
\widehat J_\eta\bigl(\hat z(G),\mathrm{onehot}(n),
E_\theta(G,\theta_\psi(G))\bigr)
+\beta\sum_{q\in\{x,y\}}
D_{\rm KL}\bigl(\theta_{\psi,q}(G)\,\|\,\theta_{\psi_0,q}(G)\bigr)
\right],\quad\rev{\beta=0.1}.
\]
\rev{Refinement uses Adam with learning rate $3\times10^{-4}$, batch size 64, at most 24 epochs and clip norm 1. Eight configurations were evaluated on DEV, each for 24 epochs: $\beta\in\{0.03,0.1,0.2,0.3,0.5\}$ with learning rate $10^{-4}$, $\beta=0.1$ with $3\times10^{-5}$ and $3\times10^{-4}$, and $\beta=0.3$ with $3\times10^{-4}$. The top three were compared on DEV2 to select one configuration shared across sizes. The selected checkpoints are epochs 19, 21 and 24 for seeds 0, 1 and 2.} The KL penalty discourages large departures from the original
initialiser, around which the surrogate was fitted. Checkpoints are selected with frozen-controller rollouts.

\paragraph{Controller observation and policy.}
The deployed actor observes a 59-dimensional vector combining frozen game
context, the present strategy, active primitive, normalised budget progress,
encoded memory and trajectory statistics. The stay/switch gate and the
conditional seven-way router are separate Tanh networks of sizes
$59\to64\to64\to2$ and $59\to64\to64\to7$, respectively. Current or
budget-infeasible primitive choices are masked. The privileged critic has
$129\to64\to64\to1$ Tanh layers and may access one-step counterfactual
primitive proposals during training only; these are not computed for the
deployed actor.

\paragraph{Controller training.}
Behavioural-cloning labels come from a rolling teacher that simulates each
of seven candidate primitives from a decision point to the remaining budget,
scoring remaining AUC plus switching cost; only the selected next window is
executed before replanning. BC uses 20 epochs, Adam $10^{-3}$, batches of
1,024 decision samples and gradient clip 1. PPO uses 1,500 updates, each
sampling 192 games (64 per size); Adam $10^{-4}$, four optimiser epochs per
batch, minibatch 512, ratio clip 0.2, entropy weight 0.01, value-loss weight
0.5 and gradient clip 0.5. Returns are undiscounted ($\gamma=1$), with GAE
$\lambda=0.95$. The first 50 updates train only the critic; checkpoints are
evaluated every 50 updates on DEV and selected by mean size-normalised AUC.
The route loss is applied only to switch choices. The training budget is
$C_{\max}=200$ operator calls, with decisions every 10 calls and Optimistic
as reference primitive. Three seeds are used.

\section{Experimental protocols and statistics}
\label{app:protocols}

\paragraph{Matched-compute evaluation.}
All iterated methods \rev{within a comparison} use the same game roots, budget and operator-call
accounting. The primary AUC uses the current strategy iterate, not the
best previously certified iterate. Policies are trained for $C=200$ calls;
$C=400$ and $C=1000$ reuse the frozen policies without retraining.
We average results across three training seeds on each root before
calculating paired differences.

\paragraph{Effect sizes and intervals.}
Let $\overline A_C^{\mathrm{M1}}$ denote the mean fixed-Optimistic AUC over
the corresponding panel. We report
\[
R_C(A)=\frac{\overline A_C(A)}{\overline A_C^{\mathrm{M1}}},\qquad
D_C(A-B)=\frac{\overline A_C(A)-\overline A_C(B)}
{\overline A_C^{\mathrm{M1}}}.
\]
Paired 95\% intervals resample independent roots: the controller-ablation table uses 2,000 bootstrap replicates, while the main paper-suite and value-guided comparisons use 10,000.
The percentage reduction of Full relative to comparator $B$ in
Table~\ref{tab:supp-feedback} is
$100[\overline A_C(B)-\overline A_C(\mathrm{Full})]/\overline A_C(B)$;
Holm correction covers the six comparisons
\{Full versus One later switch, Full versus PPO without BC\} across the three sizes
. Remaining bootstrap
intervals are pointwise unless explicitly stated otherwise.

\paragraph{Hardest 5\%.}
For each size \rev{in the 2,048-game analysis panels}, the 103 game instances (5\% of 2,048, rounded) with the largest
$C=200$ AUC under the original warm start and fixed Optimistic are marked
as the hard tail. The group is selected by the fixed-Optimistic baseline,
not by our controller.


\paragraph{Illustrative trajectories.}
For each size, examples are selected from the TEST3 hard tail defined by the
three-seed mean M1 $C=200$ AUC. Within that stratum, we determine the most
frequent Full solver-sequence pattern under seed 0 and choose the median
M1-minus-Full AUC improvement within that pattern. The examples are selected
from observed outcomes and are descriptive rather than confirmatory.

\section{Baselines and ablations}
\label{app:baselines}

\begin{table}[t]
\centering
\caption{Comparator definitions. Unless stated otherwise, learned policies
use the same primitive library and memory protocol.}
\label{tab:supp-baselines}
\footnotesize
\begin{tabular}{@{}lp{0.69\textwidth}@{}}
\toprule
Comparator & Definition \\
\midrule
M1 & Original Atlas warm start followed by fixed Optimistic; normalisation reference \\
Static selection & A single primitive selected at $C=0$ from 23 features (12-d $\hat z$, eight pooled warm-start features, three size indicators), then fixed \\
Global fixed schedule & Per size and seed, select the lowest-TRAIN-AUC of 805 plans: seven fixed primitives or one change $i\to j$ at calls $10,20,\ldots,190$; the selected plan is shared by games \\
Open-loop & Conditions on game representation and compute progress but not trajectory state \\
One later switch & Same frozen call-0 opening decision as Full, followed by a separately learned policy allowing at most one later switch. Opening changes count toward total switches and penalties \\
PPO without BC & Full policy trained with PPO but no behavioural cloning \\
BC only & Policy after cloning, with no PPO fine-tuning \\
No $\hat z$ & Full without frozen game representation input \\
Ordinary critic & Replaces training-only privileged critic with a critic observing $o_t$ \\
Restricted oracle & Searches schedules with at most three switches on the shared grid (Appendix~\ref{app:oracle}) \\
Rolling oracle & Approximate supervisor used for behavioural cloning, not an exact upper bound \\
\midrule
Full-L & One gate/router controller trained jointly at 3, 4, 5, 10, 20, 50 and 100 actions, with size-normalised observations (Appendix~\ref{app:scale-cost}) \\
TRAIN-selected fixed & The single primitive with the lowest TRAIN mean AUC for the evaluation size, start and seed, applied to every game \\
kNN selection & Per-game static selection: $k$-nearest-neighbour regression of each primitive's log AUC on $\hat z$, start features and payoff statistics; the lowest prediction is chosen, and Optimistic is replaced only when the predicted gain exceeds a margin. Hyperparameters and the margin are tuned on DEV \\
Stall switching & Every 10 calls, switch to the next primitive in the TRAIN ranking when the one-update-lagged Nash gap has fallen by less than a fraction $\delta$; the opening primitive, $\delta$ and the number of primitives cycled are tuned on DEV \\
Lightweight controller & Full-L without $\hat z$; it observes lagged-gap features computed from payoffs already evaluated by the solver \\
Four-/three-primitive & Full-L retrained with the library restricted to Optimistic, EG, fictitious play and Anchoring, or to the first three of these \\
\bottomrule
\end{tabular}
\end{table}

NfgTransformer direct predictions are measured without iterative solving;
NfgTransformer + fixed Optimistic and NfgTransformer + Full instead use
those predictions as starts for the stated iterative policies. These are
different baselines and are not treated as interchangeable.

\section{Additional results}\label{app:additional}
\subsection{Initialisation and data alignment}
\label{app:initialisation}

\paragraph{Data alignment and its limits.}
The data-matched NfgTransformer baseline (NfgT) has 285,313 parameters
(width 64, four attention heads, four blocks and feed-forward width 128). It is
trained with AdamW using learning rate $10^{-3}$, weight decay $10^{-4}$, batch
size 256, 200 warm-up steps and cosine decay. Each of three seeds completes
14,400 optimisation updates; the checkpoint is chosen by minimum DEV Nash gap
after the full training run. The best DEV gaps are 0.005480, 0.007310 and
0.007466 for seeds 0, 1 and 2.

The Atlas head and NfgT share the same 512 TRAIN and 128 DEV game instances,
corresponding to 313 and 106 independent roots. The Atlas encoder is pretrained on a 20,000-instance $3\times3$ corpus (1,068 roots), with 408 of the 512 head-training instances and all 313 head-training roots overlapping this corpus; the training objectives differ. The NfgT comparison is therefore aligned at the head
training set, not at total pretraining data, objective or training compute.
NfgT is not trained on $4\times4$ or $5\times5$ instances, making those
sizes cross-size evaluations for that initialiser.

\paragraph{\rev{Initialiser and training-data comparisons.}}
\rev{In Table~2 of the main paper, NfgT is trained on the same 512 games as the Atlas head. Trained on 12,288 games, NfgT + Full reaches $R_{200}=0.574/0.676/0.636$: comparable to Refined Atlas + Full at $3\times3$ and lower at $4\times4$ and $5\times5$. At $C=1000$ the two are indistinguishable at every size.}

\rev{Replacing the value surrogate with the log initial gap, under the same training procedure and KL regularisation, gives $R_{200}=0.556/0.688/0.666$ with Full: comparable to value-guided refinement at $3\times3$ and $5\times5$ and lower at $4\times4$.}

\begin{table}[t]
\centering
\caption{\textbf{Initial Nash gap on TEST3 before any iterative solver update.}
Smaller is better. Direct predictors are evaluated without subsequent solver
calls in this table.}
\label{tab:supp-init-gap}
\footnotesize
\setlength{\tabcolsep}{5pt}
\begin{tabular}{lccc}
\toprule
Initial strategy & $3\times3$ & $4\times4$ & $5\times5$\\
\midrule
Original Atlas warm start & 0.3008 & 0.3362 & 0.3551\\
NfgT                 & 0.0327 & 0.0835 & 0.1022\\
\bottomrule
\end{tabular}
\end{table}

\paragraph{Subgroup behaviour.}
At $C=200$ on TEST3, comparing NfgT + Optimistic minus Original Atlas +
Full gives normalised paired differences
$D=-0.409,-0.214,-0.141$ on the other 95\% of games, but
$D=+0.179,+0.152,+0.169$ on the hardest 5\%; all respective subgroup intervals
exclude zero. Thus a much stronger initial prediction is especially effective on
typical games, whereas adaptive control is more valuable in the hard tail.

Pairing NfgT with the frozen Full controller improves over Original Atlas +
Full overall, but does not show a detectable improvement in the hard 5\%:
$D=+0.009,-0.000,-0.045$ with intervals
$[-0.019,0.040]$, $[-0.041,0.037]$ and $[-0.096,0.000]$.
The gains of NfgT + Full therefore come mainly from the other 95\% of games.

\paragraph{Output-rule sensitivity.}
The main paper reports current-iterate AUC. Under the alternative
best-certified-iterate convention, NfgT + Optimistic is significantly better
than Original Atlas + Full at $C=200$ on all three sizes, with paired differences
$-0.326,-0.214,-0.205$. Current-point and certified-output results answer
different questions and are not pooled or described as interchangeable.


\begin{table}[t]
\centering
\caption{\textbf{Paired differences} $D$ (fraction of M1 AUC), 95\%
root-bootstrap intervals, pointwise. Negative favours the first method.
``Full'' is Original Atlas + Full.}
\label{tab:supp-paired-nfgt}
\footnotesize
\begin{tabular}{llccc}
\toprule
Budget & A $-$ B & $3\times3$ & $4\times4$ & $5\times5$\\
\midrule
$200$ & NfgT+Opt $-$ Full & $-0.035\,[-0.081,0.003]$ & $-0.044\,[-0.094,0.003]$ & $-0.008\,[-0.053,0.033]$\\
 & NfgT+Full $-$ Full & $-0.155\,[-0.195,-0.122]$ & $-0.088\,[-0.122,-0.056]$ & $-0.130\,[-0.161,-0.101]$\\
 & NfgT+Full $-$ NfgT+Opt & $-0.121\,[-0.147,-0.094]$ & $-0.044\,[-0.087,0.002]$ & $-0.122\,[-0.148,-0.094]$\\
\midrule
$400$ & NfgT+Opt $-$ Full & $0.017\,[-0.029,0.057]$ & $0.031\,[-0.032,0.088]$ & $0.083\,[0.032,0.131]$\\
 & NfgT+Full $-$ Full & $-0.116\,[-0.160,-0.081]$ & $-0.077\,[-0.115,-0.043]$ & $-0.110\,[-0.144,-0.080]$\\
 & NfgT+Full $-$ NfgT+Opt & $-0.133\,[-0.169,-0.097]$ & $-0.108\,[-0.157,-0.057]$ & $-0.193\,[-0.227,-0.160]$\\
\bottomrule
\end{tabular}
\end{table}

\subsection{Controller ablations}
\label{app:ablations}

\paragraph{Feedback and intervention structure.}
Under the 200-call protocol, Original Atlas + Full improves over a global fixed
schedule by 9.8\%, 11.8\%, and 12.2\% for $3\times3$, $4\times4$ and
$5\times5$, with corresponding 95\% paired root-bootstrap intervals
$[5.6,14.0]$, $[2.5,20.0]$ and $[7.3,16.9]$ percentage points.

Against an independently trained One later switch controller, Full improves by
4.5\%, 10.8\% and 9.3\%. After Holm correction over the six-test family defined in Appendix~\ref{app:protocols}, the
$4\times4$ and $5\times5$ comparisons pass ($p=0.021$ and $0.006$), whereas
$3\times3$ does not ($p=0.095$). One later switch itself remains a strong baseline,
improving over M1 by 10.7\%, 7.4\% and 10.4\%.

\paragraph{Behavioural cloning and PPO.}
PPO trained without behavioural cloning still learns to switch and improves over
M1 by 5.0\%, 8.3\% and 8.1\%. Full BC+PPO improves over No-BC by
10.2\%, 9.9\% and 11.5\%; after Holm correction, the $3\times3$ and
$5\times5$ comparisons pass ($p=0.006$ for each), while the $4\times4$
comparison does not ($p=0.064$). BC therefore improves the present
finite-budget learning procedure but is not a logical prerequisite for learning
switching behaviour. The BC-only controller is weaker than Full, supporting the
additional contribution of PPO fine-tuning.

\begin{table}[t]
\centering
\caption{\textbf{Controller ablations at $C=200$.} Relative AUC
reduction (\%) of Original Atlas + Full over each separately trained comparator.
Positive values favour Full; brackets are 95\% paired root-bootstrap intervals.}
\label{tab:supp-feedback}
\scriptsize
\setlength{\tabcolsep}{3pt}
\begin{tabular}{lccc}
\toprule
Comparator & $3\times3$ & $4\times4$ & $5\times5$\\
\midrule
Static selection
& 11.7 [6.3,17.6]
& 14.6 [6.2,21.9]
& 14.0 [6.9,20.8]\\
Open-loop (game+time)
& 4.5 [-0.2,9.0]
& 9.7 [1.3,16.9]
& 7.7 [3.4,11.9]\\
One later switch
& 4.5 [-0.7,9.2]
& 10.8 [2.9,18.0]
& 9.3 [4.5,14.0]\\
PPO without BC
& 10.2 [5.7,15.0]
& 9.9 [1.1,18.1]
& 11.5 [5.8,16.7]\\
\bottomrule
\end{tabular}
\end{table}

\paragraph{Game context, critic and intervention parameters.}
On TEST3 at $C=200$, removing $\hat z$ from the controller changes normalised AUC
by $D=+0.061,+0.032,+0.012$ relative to Full; the $5\times5$ interval contains
zero. Replacing the privileged critic with an ordinary critic changes it by only
$+0.008,+0.021,+0.017$. Both effects are smaller than that of removing trajectory feedback (static
selection in Table~\ref{tab:supp-feedback}).

Using a 5-call controller decision interval or a 5\% switching cost shows no
detectable difference from Full in the overall three-size AUC comparisons. Under
the larger switching cost, the mean switch count falls to approximately
0.9, 1.2 and 1.5 per game. The corresponding switching-cost oracle analysis
searches only the saved candidate schedules.

\paragraph{Start and controller contrasts.}
\rev{Table~\ref{tab:supp-2x2} gives the paired contrasts behind the $2\times2$ design in Table~2 of the main paper.}

\begin{table}[t]
\centering
\caption{\textbf{Paired contrasts for the $2\times2$ start--controller design}
(evaluation panel, $C=200$). Entries are AUC differences divided by M1 AUC, with
95\% paired root-bootstrap intervals; negative values favour the first-named
configuration.}
\label{tab:supp-2x2}
\footnotesize
\setlength{\tabcolsep}{3.5pt}
\begin{tabular}{lccc}
\toprule
Contrast & $3\times3$ & $4\times4$ & $5\times5$\\
\midrule
Refined vs.\ Original, fixed Optimistic & $-.263\;[-.340,-.208]$ & $-.208\;[-.260,-.163]$ & $-.232\;[-.291,-.180]$\\
Refined vs.\ Original, Full & $-.210\;[-.273,-.161]$ & $-.160\;[-.216,-.112]$ & $-.204\;[-.264,-.149]$\\
Full vs.\ fixed Optimistic, Original start & $-.228\;[-.276,-.186]$ & $-.105\;[-.169,-.033]$ & $-.122\;[-.186,-.047]$\\
Full vs.\ fixed Optimistic, Refined start & $-.176\;[-.211,-.143]$ & $-.057\;[-.122,.015]$ & $-.094\;[-.147,-.031]$\\
\bottomrule
\end{tabular}
\end{table}


\subsection{Restricted switching-oracle analysis}
\label{app:oracle}

The restricted oracle searches are exact within the specified decision grid,
the matched primitive library, persistent-memory semantics and a cap of at most
three switches. They are not unrestricted switching optima.

For the \rev{DEV2} weighted oracle subset, 403 games per size are sampled, including
all 103 games in the hardest-5\% stratum. Oracle ``headroom'' includes the
benefit of choosing the initial primitive, which is reported separately from the
incremental value of switching after that choice.

\begin{table}[t]
\centering
\caption{\textbf{Restricted oracle decomposition.} Percentages are relative to
M1 AUC or to the indicated post-initial-choice switching opportunity. \rev{Results use the weighted DEV2 subset.}}
\label{tab:supp-oracle}
\footnotesize
\setlength{\tabcolsep}{3.5pt}
\begin{tabular}{lccc}
\toprule
Quantity & $3\times3$ & $4\times4$ & $5\times5$\\
\midrule
Best initial choice, $C=200$ & 33.5\% & 38.8\% & 40.2\%\\
Additional $K\le3$ switching & 20.0\% & 20.1\% & 20.5\%\\
$K=1$ share of post-choice opportunity & 77.9\% & 68.4\% & 67.9\%\\
$K=1$ share incl.\ initial choice & 91.8\% & 89.2\% & 89.2\%\\
Full recovery of combined restricted opportunity & 38.3\% & 15.5\% & 25.4\%\\
\midrule
$K=1$ share of post-choice opportunity, $C=400$
& 78.3\% & 69.6\% & 69.0\%\\
\bottomrule
\end{tabular}
\end{table}

The rolling oracle used for behavioural-cloning supervision is an approximate
planning reference, not an upper bound. In addition, many of the hardest games
reach the three-switch cap. Their unrestricted dynamic headroom may therefore be
larger than the restricted estimates above.

\rev{Two local checks examine learned decisions directly. On 256 roots per size and three seeds, 83--96\% of sampled decisions admit no improving one-step deviation when subsequent decisions follow the frozen policy; the best of these single deviations in each game would still lower mean AUC by 0.10, 0.33 and 0.22 in units of M1 AUC at $3\times3$, $4\times4$ and $5\times5$. An exact search copies the full solver states reached by PPO at calls 170 and 180 and enumerates all 343 or 49 continuations over the remaining 30 or 20 calls. For the Full controller, the learned tail costs 21--71\% more than the optimal tail, equal to 0.4--3.0\% of the complete 200-call M1 AUC.}


\subsection{Frozen-representation diagnostics}
\label{app:representation}

The frozen behavioural representation contains some information about whether
adaptive control is likely to be useful, but much less about the magnitude or
timing of that benefit. On DEV2, a small probe predicting whether beneficial
switching opportunity exists obtains AUROC 0.738 (95\% CI
$[0.707,0.766]$). Predicting the magnitude of switching headroom is weak
($R^2=0.051$ overall), while predicting the first optimal switch time is weaker
still ($R^2=-0.021$, Spearman $0.059$).

These probes support the architectural
separation used in the paper: $\hat z(G)$ is a persistent game-level prior, while
the online optimisation state supplies the information needed for the exact
switching decision.

The end-to-end controller ablation is consistent with this interpretation:
removing $\hat z$ increases normalised AUC by
$0.061,0.032,0.012$ on $3\times3$, $4\times4$ and $5\times5$ respectively,
with no detected effect at $5\times5$.


\subsection{Certification, compute-to-target and runtime}
\label{app:compute}

\paragraph{Operator-call accounting.}
We count payoff/operator evaluations rather than nominal iterations:
extragradient costs two evaluations per update and the other implemented
primitives one. Current-iterate AUC is the primary statistic in the main paper.

For certified $\varepsilon$-success, the initial profile must be queried at the
stated one-call certification cost, and a state that was not queried at the end
of the budget cannot be treated as certified. Success fractions and calls to
$\varepsilon$ must be read jointly, because mean hitting time conditional on
success can otherwise favour a method that fails on difficult games.

\begin{table}[t]
\centering
\caption{\textbf{Certified $\varepsilon=10^{-3}$ results on TEST3 at $C=200$.}
Mean calls are conditional on certified success and must be interpreted together
with the success fractions.}
\label{tab:supp-certified}
\footnotesize
\setlength{\tabcolsep}{3.5pt}
\begin{tabular}{lcccccc}
\toprule
& \multicolumn{3}{c}{Success (\%)} & \multicolumn{3}{c}{Mean calls}\\
Method & $3\times3$ & $4\times4$ & $5\times5$
& $3\times3$ & $4\times4$ & $5\times5$\\
\midrule
M1 & 94.3 & 88.5 & 81.7 & 25.9 & 51.9 & 69.3\\
Original Atlas + Full & 96.1 & 91.7 & 86.0 & 17.8 & 42.9 & 60.4\\
NfgT + Full & 96.0 & 91.7 & 87.0 & 14.3 & 37.3 & 50.3\\
\bottomrule
\end{tabular}
\end{table}

\paragraph{Controller overhead.}
At batch size one on the measured CPU configuration, the controller forward pass
takes approximately 0.23, 0.22 and 0.24 ms per decision for the three game sizes.
At $C=200$, measured end-to-end Full/M1 runtime ratios are 1.18, 1.09 and 0.98.
At batch size 64, Full is approximately 1.6--1.9 times slower than M1.
\rev{Appendix~\ref{app:scale-cost} reports CPU/GPU elapsed-time comparisons and a measured-unit-cost analysis.}


\section{Pseudocode}
\label{app:pseudocode}
\begin{algorithm}[t]
\caption{Deterministic deployment of Full: one decision per 10-call window}
\label{alg:solver}
\begin{algorithmic}[1]
\REQUIRE Game $G$; frozen recogniser $f_\phi$ and refined initialiser $I_\psi$;
 gate $\pi_{\rm gate}$ and router $\pi_{\rm route}$; primitives
 $\{\mathcal T_j\}_{j=1}^{M}$; budget $C_{\max}$;
 $\Delta_{\rm dec}=10$
\STATE $\hat z\gets f_\phi(G)$;
 $(H_x,H_y)\gets\mathrm{ActionEncoder}(G)$;
 $\theta\gets I_\psi(H_x,H_y,\hat z)$
\STATE $\mathcal M\gets\{\mu^{(1)},\ldots,\mu^{(M)}\}$; initialise global memory and trajectory history $h$
\STATE $i\gets\mathrm{Optimistic}$; $C\gets0$
\STATE Define $\kappa(i)=2$ for Extragradient, and $\kappa(i)=1$ otherwise
\WHILE{$C<C_{\max}$}
  \STATE $o\gets\mathrm{Observe}(\hat z,\theta,i,C/C_{\max},\mathcal M,h)$
  \STATE $g\gets\arg\max\pi_{\rm gate}(\cdot\mid o)$
  \IF{$g=\textsc{Switch}$}
    \STATE $j\gets\arg\max\pi_{\rm route}(\cdot\mid o,i)$, masking current and infeasible primitives
    \STATE $i\gets j$; record switch (including at $C=0$)
  \ENDIF
  \STATE $C_{\rm end}\gets\min(C+\Delta_{\rm dec},C_{\max})$
  \WHILE{$C+\kappa(i)\le C_{\rm end}$}
    \STATE $(\theta,\mu^{(i)})\gets\mathcal T_i(G,\theta,\mu^{(i)})$
    \STATE $C\gets C+\kappa(i)$; update global memory and history $h$
  \ENDWHILE
  \IF{$C<C_{\rm end}$}
    \STATE \textbf{break} \COMMENT{No full update fits in the remaining window}
  \ENDIF
\ENDWHILE
\RETURN Final strategy and gap, call and switch trajectories
\end{algorithmic}
\end{algorithm}

The first decision occurs at $C=0$. A change away from Optimistic at the
opening is subject to the gate and counts in the switching-cost objective.
This differs from the free-initial-selection ablation, which chooses the
initial primitive directly without the stay/switch gate. Deployment uses
argmax decisions, not stochastic sampling; privileged critic inputs are
training-only.

\section{Reproducibility}
\label{app:evidence}

All comparisons use specified root-level splits, fixed seeds, frozen checkpoints,
persistent per-primitive memory, explicit primitive-action masks and operator-call
accounting. Paired bootstrap intervals are computed after first averaging the
three training seeds on each root. They quantify root-to-root sampling variability
conditional on the trained policies.

TEST3 had already been opened during method development and is described as such.
\rev{The ablation and evaluation panels were generated after the models they evaluate were frozen; root and payoff-signature de-duplication separates the evaluation panel from development data. Table~\ref{tab:supp-panels} lists the panels.}

NfgT's $4\times4$ and $5\times5$ evaluations are cross-size only for the initialiser,
because the Full controller itself is trained jointly across all three sizes.


\section{\rev{Size transfer and computational performance}}
\label{app:scale-cost}

\subsection{\rev{Transfer of the switching policy}}
Full-L is trained jointly on games with 3, 4, 5, 10, 20, 50 and 100 actions; sizes 15, 30 and 75 are held out and test interpolation within the training range. Table~\ref{tab:supp-scale-all} compares it with three baselines at uniform starts and $C=200$ on the multi-size evaluation panel (512 games per size; 1,024 at 50 and 100 actions). The TRAIN-selected primitive is fictitious play at 3--5 actions, EG at 10 and Anchoring at 15--100. Full-L improves on all three baselines at sizes 15 and 30 and on stall switching at 75, and its mean AUC is lower than that of every fixed primitive at every size, by 2--13\%.

\begin{table*}[t]
\centering
\caption{Size-transfer comparisons at the training sizes and the held-out sizes 15, 30 and 75, with uniform starts and $C=200$. Entries are $(\overline{\mathrm{AUC}}_{\mathrm{Full-L}}-\overline{\mathrm{AUC}}_B)/\overline{\mathrm{AUC}}_{\mathrm{M1}}$, with paired 95\% intervals. M1 is Original Atlas + Optimistic on the same panel, size and budget. Negative values favour Full-L. A dagger marks $p<0.05$ after Holm correction across all 30 tests; positive significant entries are losses. Held-out sizes are marked with an asterisk.}
\label{tab:supp-scale-all}
\small
\begin{tabular}{rccc}
\toprule
Actions & TRAIN-selected fixed & kNN selection & Stall switching \\
\midrule
3 & $-.943\;[-1.419,-.647]^\dagger$ & $.096\;[.052,.160]^\dagger$ & $-.031\;[-.115,.047]$ \\
4 & $-1.065\;[-1.394,-.810]^\dagger$ & $-.068\;[-.130,-.009]$ & $-.309\;[-.404,-.222]^\dagger$ \\
5 & $-.785\;[-1.003,-.602]^\dagger$ & $-.075\;[-.133,-.019]$ & $-.173\;[-.239,-.112]^\dagger$ \\
10 & $-.172\;[-.231,-.112]^\dagger$ & $-.096\;[-.139,-.054]^\dagger$ & $-.049\;[-.098,-.0001]$ \\
15* & $-.129\;[-.179,-.084]^\dagger$ & $-.085\;[-.127,-.045]^\dagger$ & $-.154\;[-.201,-.109]^\dagger$ \\
20 & $-.079\;[-.111,-.048]^\dagger$ & $-.065\;[-.099,-.030]^\dagger$ & $-.091\;[-.124,-.056]^\dagger$ \\
30* & $-.039\;[-.062,-.016]^\dagger$ & $-.046\;[-.075,-.017]^\dagger$ & $-.225\;[-.268,-.183]^\dagger$ \\
50 & $-.019\;[-.035,-.003]$ & $.001\;[-.015,.018]$ & $-.097\;[-.120,-.075]^\dagger$ \\
75* & $-.021\;[-.043,.001]$ & $.001\;[-.020,.021]$ & $-.089\;[-.116,-.061]^\dagger$ \\
100 & $-.024\;[-.032,-.015]^\dagger$ & $.018\;[.008,.029]^\dagger$ & $-.053\;[-.067,-.039]^\dagger$ \\
\bottomrule
\end{tabular}
\end{table*}

\subsection{\rev{Operator-call efficiency and elapsed time}}
Lower operator-call AUC indicates more effective use of payoff evaluations. To assess computational performance, we measure elapsed-time quality with representation inference and controller decisions included, and use a unit-cost model to identify when call savings compensate for overhead.

The CPU evaluation uses the timing subset (Table~\ref{tab:supp-panels}) on an Apple M4 Pro, with one process and one thread. It compares the lightweight controller and Full-L with fixed Optimistic, EG and Anchoring, kNN selection and stall switching, using the time-averaged gap: the mean Nash gap from time zero to each deadline. Each method--game run is repeated three times in interleaved order, taking the median before averaging model seeds. The time budgets are calibrated on DEV as the time for fixed Optimistic to execute 200, 1,000 and 5,000 calls. At 100 actions these are 1.96, 9.44 and 45.76 ms. Representation inference, observation construction, controller inference, switching, memory maintenance and solver updates are charged. Evaluation gaps are computed offline; online stopping certification is not included.

The lightweight controller removes the behavioural representation and uses the previous update's gap, derived from payoff vectors already computed by that update. Its arithmetic is timed, although it requires no additional operator evaluation. At 100 actions it reduces time-averaged gap by approximately 23\% and 27\% relative to Optimistic at the first two deadlines, and by approximately 7\% relative to stall switching, with paired 95\% intervals excluding zero. These results demonstrate a regime in which learned decisions repay their overhead. At 100 actions the lightweight controller also improves on EG at the first deadline and on Anchoring at the second and third; it ties EG at the second deadline, and EG is better at the third. At 20 actions and the first deadline it is worse than every fixed solver and stall switching. Full-L, which also pays for representation inference (1.6, 8.6 and 37 ms at 20, 50 and 100 actions), has a higher time-averaged gap than Optimistic, EG, Anchoring and stall switching at every size and deadline.

We repeat the evaluation on an AMD EPYC 7R32 CPU with an NVIDIA A10G GPU, which computes the representation and the start; the solver and controller run on one CPU thread. Representation and start inference are batched over the 256 games of each size, and each game is charged 1/256 of the batch time. At 100 actions, Full-L with the Original Atlas start is significantly better in 3 and worse in 7 of the 10 Holm-corrected comparisons with five baselines at the first two deadlines: it improves on Optimistic at the second deadline and on GPU-based kNN at both, and loses to EG, Anchoring and stall switching at both.

\paragraph{\rev{Training and amortisation.}}
The small-game pipeline cost ledger totals approximately 3.0 CPU core-hours, dominated by teacher labels (42\%) and PPO (33\%). This total combines recorded costs with estimates for recogniser training and behavioural cloning. A measured-unit-cost model predicts that switching alone repays decision overhead at roughly 3--10 $\mu$s per operator call in the 3--5-action setting; representation and initialisation raise the full-pipeline threshold to roughly 0.1--0.3 ms per call. Relative to uniform-start Optimistic run to a certified gap of $10^{-3}$, the model predicts that the 3.0 core-hours are repaid after 46,000--61,000 solves at an assumed 10 ms per operator call. Training Full-L for three seeds takes 23.76 core-hours: 15.64 for behavioural-cloning labels and 8.12 for cloning and PPO.

\subsection{\rev{Terminal accuracy and the objective being optimised}}
AUC and terminal accuracy can rank methods differently. On the multi-size evaluation panel (uniform starts, $C=200$), Full-L reaches gap $\le0.01$ on 50\%, 26\% and 16\% of games at 20, 50 and 100 actions, against 78\%, 62\% and 49\% for EG, although its AUC is lower at each of these sizes.

\section{\rev{Primitive-portfolio ablation}}
\label{app:portfolio}

We retrain Full-L with a four-primitive library (Optimistic, EG, fictitious play and Anchoring) and a three-primitive library (without Anchoring), three seeds each, and compare them across 60 conditions on the multi-size evaluation panel (ten sizes, two starts and three budgets). After Holm correction over the 60 conditions, the four-primitive controller is better than Full-L in 11 conditions, all at 30--100 actions, and worse in none; Full-L is better than the three-primitive controller in 17 and worse in none. Four primitives therefore suffice at these sizes, and removing Anchoring as well loses performance.

\section*{AI Usage Statement}

AI-assisted tools were used to support language
polishing, LaTeX formatting, and code development.
All scientific contributions, experimental results,
and conclusions were reviewed and validated by
the authors, who take full responsibility for
the content of this paper.

\end{document}